\documentstyle[12pt,epsfig,amsmath]{article}
\begin{document}

\begin{center}

{\Large FAMILIES OF COMPLEX HADAMARD MATRICES}

\vspace{18mm}

{\large Nuno Barros e S\'a}\footnote{nunosa@uac.pt}

\vspace{6mm}

{\sl DCTD, Universidade dos A\c{c}ores,}

{\sl 9500-801 Ponta Delgada, Portugal}

\vspace{10mm}

{\large Ingemar Bengtsson}\footnote{ibeng@fysik.su.se}

\vspace{6mm}

{\sl Fysikum, Stockholms Universitet,}

{\sl S-106 91 Stockholm, Sweden}

\vspace{20mm}

{\bf Abstract:}

\end{center}

\

\noindent What is the dimension of a smooth family of complex
Hadamard matrices including the Fourier matrix? We address this problem with
a power series expansion. Studying all dimensions up to 100 we find that the
first order result is misleading unless the dimension is 6, or a power of a
prime. In general the answer depends critically on the prime number
decomposition of the dimension. Our results suggest that a general theory
is possible. We discuss the case of dimension 12 in
detail, and argue that the solution consists of two 13-dimensional
families intersecting in a previously known 9-dimensional family.
A precise conjecture for all dimensions equal to a prime times another
prime squared is formulated.

\newpage

{\bf 1. Introduction}

\hspace{5mm}

\noindent A type of mathematical problem that arises very often is that one
has a set of $M$ algebraic equations in $N$ variables,

\begin{equation} f_m(x_1, \dots , x_N) = 0 \ , \hspace{8mm} m = 1, \dots , M \ .
\label{1} \end{equation}

\noindent One solution is known, and one asks whether this is an isolated
solution, and if not one asks for the dimension of the solution space around
that special solution. The problem to be discussed here is of that type, and
our aim is to perform a perturbative expansion around the special solution,
in order to learn something about the possible answers to the above questions.
We will outline a systematic and completely general method to do so.

The particular problem that we will study concerns complex Hadamard
matrices. By definition a complex Hadamard matrix is a unitary $N\times N$
matrix all of whose entries have the same modulus.
The classification of such matrices has a long history \cite{Sylvester}.
It has been completed for $N \leq 5$ \cite{Haagerup}, and much partial
information is available in higher dimensions \cite{TZ, Horadam, Feriphd}. This
classification problem is of interest for many reasons. In quantum theory it
is equivalent to the problem of classifying all complementary pairs of bases
\cite{Kraus}---now known as Mutually Unbiased Bases \cite{Durt}---and it can be used
to give quantitative content to Bohr's principle of complementarity \cite{Berge}.
It also forms part of the problem of classifying unitary operator bases \cite{Werner},
which is of importance in quantum information theory, and it arises in operator
algebra \cite{Popa}, quantum groups \cite{Banica}, and elsewhere.

A complex Hadamard matrix that exists for any $N$ is the Fourier matrix, whose
entries are

\begin{equation} F_{ij} = \frac{\omega^{ij}}{\sqrt{N}} \ , \hspace{8mm} 0 \leq i,j \leq N-1
\ , \hspace{8mm} \omega = e^{\frac{2\pi i}{N}} \ . \end{equation}

\noindent This is the special solution that we are going to expand around.
Although we will proceed somewhat differently, in principle one can proceed
with this problem by modifying all the matrix elements according to

\begin{equation} \omega^{ij} \rightarrow \omega^{ij}e^{i\tau_{ij}} \ ,
\end{equation}

\noindent expanding the phase factors in powers of $\tau_{ij}$, and then solving
the unitarity equations order by order in $\tau_{ij}$. To first order in
the perturbation this has been done by Tadej and \.Zyczkowski \cite{ZT},
who found that the number of free parameters that survive to first order is

\begin{equation} D_1 = \sum_{n=0}^{N-1} \mbox{gcd}(n,N) \ , \label{TaZy}
\end{equation}

\noindent where gcd denotes the greatest common divisor. If $N$ is prime
this means that the number of free parameters is $2N-1$. This is the number
one would naively expect by counting the number of equations and the number
of parameters in the matrix. However, for non-prime values of
$N$ the number $D_1$ is larger than that.

We can multiply the rows and columns of a complex Hadamard matrix with $2N-1$
overall phase factors, while still preserving the Hadamard property. Hence $2N-1$
parameters are trivial, and it is customary to remove them by presenting
complex Hadamard matrices in dephased form, such that the first row and the first
column have real and positive entries. The number

\begin{equation} d_1 = D_1 - (2N-1) = \sum_{n=1}^{N-1} (\mbox{gcd}(n,N) - 1)
\label{defect} \end{equation}

\noindent has been called the defect of the Fourier matrix
\cite{TZ}, and gives an upper bound on the dimension of any smooth
family of dephased complex Hadamard matrices passing through the
Fourier matrix. Here we shall call it the linear defect since we
found a way of defining order by order a number that is a natural
generalization to higher orders of the defect and that coincides
with it to first order; we shall then speak of the nth order defect.
If $N = p^k$ is a prime power the situation is very satisfactory
since explicit solutions for dephased Hadamard matrices with $d_1$
free parameters are known. These solutions are of an especially
simple kind, known as affine families \cite{TZ}. We are concerned
with other values of $N$, for which very little is known. An
exception is the special case $N = 6$, for which $d_1 = 4$, a
3-dimensional family has been constructed explicitly
\cite{Karlsson}, and a 4-dimensional family has been shown to exist
\cite{Szollosi, Feriphd}. What is missing so far is a proof that the
4-dimensional family includes the Fourier matrix, and that it is
smooth there. Numerical arguments for the truth of this have been
presented \cite{Skinner}, and we will confirm this conclusion in a
perturbative expansion that we carried to order 100.

But it appears that the first order result is misleading in all other
cases. Indeed for all values of $N$ less than 101 and not equal to 6
or to a prime power, we will prove that the dimension of the largest
smooth family is strictly less than the linear defect.

What we really want to know is by how much the dimension of the solution
space differs from $d_1$. For $N = 12$, which is of some special interest
\cite{TZ}, we find that the largest smooth family has dimension
13, while the linear defect is 17 and the largest explicitly known families have
dimension 9. The solution in fact consists of two 13-dimensional families
intersecting in a 9-dimensional affine family. Our evidence suggests that
the situation is similar whenever $N$ is a product of three primes, two of
them equal, and we are able to formulate an appealing conjecture
of what the final picture is likely to be in this case.

While these results fall short of providing a general solution to our problem
they do make us feel that a general solution of a reasonably simple sort does
exist.

Most of the paper is devoted to explaining the procedure that leads to
the results we have. In section 2 we give a bird's eyes view of the general method.
A linear system of equations has to be solved at each order of the expansion,
but the problem is non-trivial because it can happen that a
solution exists only if non-linear
consistency conditions are imposed on the lower order
solutions. In section 3 we begin our discussion of complex Hadamard matrices,
and set up the equations that are to be solved perturbatively. In sections 4, 5,
and 6 we describe how the general method applies to the problem at hand. In section
7 we investigate all $N \leq 100$, and find out when the dimension
of any smooth family is indeed less than $d_1$. In section 8 we discuss the
case of $N = 12$ in full detail, and in section 9 we discuss the cases $N = 18, 20$,
ending with a conjecture for all $N = p_1p_2^2$, where $p_1, p_2$ are prime
numbers.

The calculations were partly numerical, partly symbolical, and were performed
using Mathematica. We warn the reader at the outset that our results rely
on a considerable
amount of calculational detail. Depending on interests the reader may therefore
want to just glance at section 2---and perhaps at the toy model in Appendix D---and 
then go directly to section 10 where
our conclusions are summarised (and where we offer some speculations
concerning the final picture).

In Appendix A we give some background
information concerning complex Hadamard matrices, we define the notion of affine families, and we prove some results concerning them that we need in the main text.
Appendix B gives an explicit construction of affine families
for $N$ equal to a prime power. Appendix C contains an exact
second order calculation, valid for all $N$. 

For ease of expression we refer to complex Hadamard matrices simply as
``Hadamard matrices'' from now on, and when
we talk of ``the dimension of the solution space'' we refer to the dimension
of the set of dephased Hadamard matrices connected to the Fourier matrix.

\vspace{10mm}

{\bf 2. The method}

\vspace{5mm}

\noindent Let us return to eqs. (\ref{1}). It is assumed that we know one
solution for the $N$ variables. By shifting the variables if necessary,
we can arrange that $x_n = 0$ is a solution of the equations. When
we expand the system around this solution the equations take the form

\begin{equation} f_m = A_{m;n}x_n + A_{m;n_1n_2}x_{n_1}x_{n_2} + ... = 0
\ . \label{2} \end{equation}

\noindent Summation over repeated indices is understood. Assume that the
solution $x_n = 0$ belongs to a family of solutions $x_n(\tau_1, \dots , \tau_D)$
with $D$ unknown parameters. If the variables are expanded in these parameters
we will obtain an expansion of the form

\begin{equation} x_n = x^{(1)}_n + x^{(2)}_n + x^{(3)}_n + ... \ . \label{expand} \end{equation}

\noindent We do not explicitly write out the dependence on the parameters.
Inserting this into eqs. (\ref{2}) we obtain

\begin{equation} f_m = A_{m;n}x^{(1)}_n + A_{m;n}x^{(2)}_n + A_{m;n_1n_2}
x^{(1)}_{n_1}x^{(1)}_{n_2} + ... = 0 \ . \end{equation}

\noindent This equation is now to be solved order by order in the
hidden parameters. Thus we obtain

\begin{eqnarray} A_{m;n}x_n^{(1)} = 0 \hspace{90mm} \\
\nonumber \\
A_{m;n}x_n^{(2)} = - A_{m;n_1n_2}x_{n_1}^{(1)}x_{n_2}^{(1)} \hspace{63mm} \\
\nonumber \\
A_{m;n}x_n^{(3)} = - A_{m;n_1n_2}(x^{(2)}_{n_1}x^{(1)}_{n_2} + x^{(1)}_{n_1}x^{(2)}_{n_2})
- A_{m;n_1n_2n_3}x^{(1)}_{n_1}x^{(1)}_{n_2}x^{(1)}_{n_3} \hspace{2mm} \end{eqnarray}

\begin{eqnarray}
A_{m;n}x^{(4)}_n = - A_{m;n_1n_2}(x^{(3)}_{n_1}x^{(1)}_{n_2} + x^{(1)}_{n_1}x^{(3)}_{n_2}
+ x_{n_1}^{(2)}x_{n_2}^{(2)}) \hspace{25mm} \nonumber \\ \nonumber \\
-A_{m;n_1n_2n_3}(x_{n_1}^{(2)}x_{n_2}^{(1)}x_{n_3}^{(1)} +
x_{n_1}^{(1)}x_{n_2}^{(2)}x_{n_3}^{(1)} + x_{n_1}^{(1)}x_{n_2}^{(1)}x_{n_3}^{(2)})
 \hspace{3mm}  \\ \nonumber \\
-A_{m;n_1n_2n_3n_4}x_{n_1}^{(1)}x_{n_2}^{(1)}x_{n_3}^{(1)}x_{n_4}^{(1)} \ ,
\hspace{43mm} \nonumber \end{eqnarray}

%
%

\noindent and so on to higher orders. At each order then we are faced with
a linear system; homogeneous to first order and heterogeneous to the higher
orders. At each order higher than the first this linear system may or may
not---depending on the form of the original equations---impose restrictions
on the solutions obtained in lower orders.

To deal with these linear systems we introduce a pseudo-inverse
of the $M\times N$ matrix $A_{m;n}$. This is a matrix $\hat{A}$ obeying
$A\hat{A}A = A$. For this general discussion it is natural
to choose the Moore-Penrose inverse for this purpose, since---due to some extra
conditions---it is
uniquely defined and easily computed \cite{Penrose}. The relevant theorem
is that the linear system

\begin{equation} AX = B \end{equation}

\noindent admits a solution if and only if

\begin{equation} ({\bf 1} - A\hat{A})B = 0 \ . \end{equation}

\noindent When this consistency condition is obeyed the general solution is

\begin{equation} X = \hat{A}B + ({\bf 1} - \hat{A}A)Z \ , \end{equation}

\noindent where the $N$ components of the column vector $Z$ are arbitrary.
The matrix $ {\bf 1} - \hat{A}A$ typically has rank less than $N$, so
the number of free parameters in the homogeneous solution is typically
less than $N$.

The first order equations are homogeneous, and admit the homogeneous solution

\begin{equation} x_n^{(1)} = (\delta_{nn_1}-\hat{A}_{n;m}A_{m;n_1})z^{(1)}_{n_1}
\equiv h_n^{(1)} \ . \end{equation}

\noindent The number of free parameters in the solution gives an upper bound
on the dimension of the solution space. To second order the solution is a
sum of a homogeneous and a heterogeneous part,

\begin{equation} x_n^{(2)} = h_n^{(2)} - \hat{A}_{n;m}A_{m;n_1n_2}h_{n_1}^{(1)}
h_{n_2}^{(1)} \ , \end{equation}

\noindent provided that the consistency condition

\begin{equation} (\delta_{mm_1}- A_{m;n}\hat{A}_{n;m_1})A_{m_1;n_1n_2}h_{n_1}^{(1)}
h_{n_2}^{(1)} = 0 \end{equation}

\noindent holds. Assume for the sake of the argument that it does, and write the
solution as

\begin{equation} x_n^{(2)} = h_n^{(2)} + H_n^{(2)} \ , \end{equation}

\noindent where the heterogeneous part is a known second order polynomial in
$h_n^{(1)}$ (obeying $h_n^{(1)} \rightarrow \lambda h_n^{(1)} \Rightarrow 
H_n^{(2)} \rightarrow \lambda^2H_n^{(2)}$). To third order the heterogeneous 
part of the solution is

\begin{equation} H_n^{(3)} = - \hat{A}_{n;m}A_{m;n_1n_2}(x^{(2)}_{n_1}x^{(1)}_{n_2}
+ x_{n_1}^{(1)}x_{n_2}^{(2)}) - \hat{A}_{n;m}A_{m;n_1n_2n_3}x^{(1)}_{n_1}x_{n_2}^{(1)}
x_{n_3}^{(1)} \ , \end{equation}

\noindent provided that

\begin{eqnarray} ({\bf 1} - A\hat{A})_{m,m_1}A_{m_1;n_1n_2}(x^{(2)}_{n_1}x^{(1)}_{n_2}
+ x_{n_1}^{(1)}x_{n_2}^{(2)}) + \nonumber \\ \ \\
+ ({\bf 1} - A\hat{A})_{m,m_1}A_{m_1;n_1n_2n_3}x^{(1)}_{n_1}x^{(1)}_{n_2}x_{n_3}^{(1)}
= 0 \ . \nonumber \end{eqnarray}

\noindent Due to the symmetrisation, and the second order consistency condition,
this condition can be written as

\begin{equation} ({\bf 1} - A\hat{A})_{m,m_1}\left( A_{m_1;n_1n_2}(H^{(2)}_{n_1}
h^{(1)}_{n_2} + h_{n_1}^{(1)}H_{n_2}^{(2)})
+ A_{m_1;n_1n_2n_3}h^{(1)}_{n_1}h^{(1)}_{n_2}h_{n_3}^{(1)} \right)
= 0 \ . \end{equation}

\noindent This is a third order polynomial in $h_n^{(1)}$. As long as the consistency
conditions continue to hold to order $s-1$, the consistency condition at order $s$
will always be a set of polynomial equations of order $s$ in $h_n^{(1)}$. If they
hold to all orders the homogeneous solutions of order higher than one can be set
to zero by means of a redefinition of the first order solution, and we are done.

Assume for the sake of the argument that the consistency conditions do break down
at third order. Then one must solve a set of third order polynomial equations
that will
restrict the first order homogeneous solution. We cannot say by how much the
upper bound on the dimension
of the solution space drops until these equations have been solved.

When we continue to the next order we can still write

\begin{equation} x_n^{(4)} = h_n^{(4)} + H_n^{(4)} \ , \end{equation}

\noindent where the heterogeneous part is a polynomial of fourth
order in the parameters and depending on $h_n^{(1)}$, $h_n^{(2)}$, 
and $h_n^{(3)}$. The consistency condition at fourth order is

\begin{eqnarray} ({\bf 1} - A\hat{A})_{mm_1}\big( A_{m_1;n_1n_2}(H^{(3)}_{n_1}
h^{(1)}_{n_2} + h^{(1)}_{n_1}H^{(3)}_{n_2} + \hspace{30mm} \nonumber \\ \nonumber \\
+ h^{(2)}_{n_1}H^{(2)}_{n_2} +
H^{(2)}_{n_1}h^{(2)}_{n_2} + H^{(2)}_{n_1}H^{(2)}_{n_2}) + \hspace{15mm}  \nonumber \\
\nonumber \\
+ A_{m_1;n_1n_2n_3}(h^{(2)}_{n_1}h^{(1)}_{n_2}h^{(1)}_{n_3} + h^{(1)}_{n_1}
h^{(2)}_{n_2}h^{(1)}_{n_3} + h^{(1)}_{n_1}h^{(1)}_{n_2}h^{(2)}_{n_3} + \nonumber \\
\ \\
+ H^{(2)}_{n_1}h^{(1)}_{n_2}h^{(1)}_{n_3} + h^{(1)}_{n_1}H^{(2)}_{n_2}
h^{(1)}_{n_3} + h^{(1)}_{n_1}h^{(1)}_{n_2}H^{(2)}_{n_3}) \hspace{5mm} \nonumber  \\
\nonumber \\
+ A_{m_1;n_1n_2n_3n_4}h^{(1)}_{n_1}h^{(1)}_{n_2}h^{(1)}_{n_3}h^{(1)}_{n_4}\big) = 0
\ . \hspace{15mm} \nonumber \end{eqnarray}

\noindent Note that $h^{(3)}_n$ does not appear because the
consistency condition is fullfilled at order two. This is a set of
linear equations for the $h^{(2)}_n$, briefly $Uh^{(2)}_n=V$. The solution 
will be a quotient of two polynomials in $h^{(1)}_n$ which is homogeneous 
of order two in that variable. If the rank of 
the matrix $U$ is
equal to the number $u$ of variables $h^{(1)}_n$ fixed by the third
order consistency equations then the corresponding variables are
fixed once and for all (it is easy to check that to fifth order one
gets consistency equations of the type $Uh^{(3)}_n=W$, where 
$U$ is the same matrix). However these linear systems come with
their own consistency conditions which, if not automatically
satisfied, will further fix the $h^{(1)}_n$ and further lower the
defect.

We have not investigated in any systematic manner what happens if
these higher order consistency conditions break down at some higher
order--although this does happen in some of the concrete cases that
we study below. If the rank of $U$ is lower than $u$ then the
remaining $h^{(2)}_n$ must be fixed at higher order by non-linear
equations (as would be the case for systems having sets of
solutions which are tangent at the point around which we are
expanding). This did not happen in the particular cases under study; 
also the toy model in Appendix D provides a reassuring consistency 
check.

It is time to turn to the subject we want to deal with.
We will then be able to adapt our method to the special form of the equations
we encounter. In particular the Moore-Penrose inverse will not be needed.

\vspace{10mm}

{\bf 3. The Hadamard equations}

\vspace{5mm}

\noindent The equations ensuring that a matrix is a Hadamard matrix
define an algebraic variety of some sort \cite{Dita}. Since our aim is to
solve these equations order by order in an expansion around the Fourier matrix
$F$ we begin by performing a discrete Fourier transformation of all our
matrices, that is

\begin{equation} H \rightarrow M = HF^\dagger \ , \end{equation}

\noindent where the dagger denotes hermitian conjugation. This means that
the Fourier matrix itself is represented by the
unit matrix. We prefer to represent the Fourier matrix with the zero matrix,
so we shift the matrix to

\begin{equation} X = {\bf 1} - M \ . \end{equation}

\noindent We must now formulate the condition that $H$ be a Hadamard
matrix as a set of equations for the matrix $X$. In order to be able to
address the equations in an efficient way we aim for a subset
of equations in which the complex conjugates of the matrix elements do
not occur.

For this purpose we introduce the permutation matrix $P$ with matrix
elements

\begin{equation} P_{ij} = \delta_{i+1,j} \ . \end{equation}

\noindent This matrix generates the cyclic group of order $N$, and has
the columns of the Fourier matrix as its eigenvectors. We denote the
commutator of two matrices by $[A,B]$, and the vector of diagonal elements
of a matrix by diag$(A)$.

\

\noindent \underline{Theorem 1}: {\sl $H$ is a Hadamard matrix if and only if
$M = HF^\dagger$ obeys}

\begin{equation} MM^\dagger = {\bf 1} \end{equation}

\begin{equation} \mbox{diag}(MP^nM^\dagger) = 0 \ , \hspace{8mm} n \neq 0 \
\mbox{mod} \ N \ ,
\label{cond2} \end{equation}

\noindent {\sl where $P$ is the permutation matrix defined above. For the
matrix $X = {\bf 1} - M$ these equations read}

\begin{equation} X + X^\dagger = XX^\dagger \label{ETT} \end{equation}

\begin{equation} \mbox{diag} \left( \sum_{p=0}^\infty [P^n,X]X^p \right)
= 0 \ , \hspace{8mm} n \neq 0 \ \mbox{mod} \ N \ ,  \label{TVA} \end{equation}

\noindent {\sl at least in a neighbourhood of the Fourier matrix.}

\

\noindent \underline{Proof}: Direct calculation shows that

\begin{equation} (MP^nM^\dagger )_{ii} = \sum_j\omega^{nj}|H_{ij}|^2
\ , \end{equation}

\noindent and elementary properties of the discrete Fourier transform
now establish that eq. (\ref{cond2}) is
necessary and sufficient, given that $M$ is unitary. Since unitarity
is imposed separately we may replace the matrix $M^\dagger$ by the
matrix $M^{-1}$ in eq. (\ref{cond2}). Expanding $M^{-1} = ({\bf 1}-X)^{-1}$
in a geometric series and using the property that

\begin{equation} \mbox{diag}\left( P^n\sum_{p=0}^\infty X^p\right) =
\mbox{diag} \left( P^n\sum_{p=0}^\infty X^{p+1} \right) \end{equation}

\noindent we arrive at eq. (\ref{TVA}). The unitarity condition on $M$
is clearly equivalent to eq. (\ref{ETT}). $\Box$

\

The plan now is to first solve eq. (\ref{TVA}) order by order in a perturbative 
expansion, using complex entries with no complex conjugates appearing, 
and to impose the unitarity condition at the end. The alternative, to impose 
unitarity at each order from the beginning, is less convenient for symbolic 
calculations using Mathematica. 

To deal with eq. (\ref{TVA}) following the method outlined in section 2 the 
first step is to expand $X$ order by order in the unknown parameters. To follow 
the letter of the method we should reshape the matrix $X$ into a vector, but 
we will not actually do so since it will turn out that we can solve directly 
for the matrix $X$. 

Thus we write 

\begin{equation} X = \sum_{s=1}^\infty X^{(s)} \ . \label{expansion} \end{equation}

\noindent Inserting this into eq. (\ref{TVA}), which itself contains an
infinite sum to be expanded, we find that the equations
at each order $s$ form the inhomogeneous linear system

\begin{equation} \mbox{diag}\left( [P^n,X^{(s)}] + B^{(s,n)} \right) = 0 \ ,
\label{higher} \end{equation}

\noindent where $B^{(s,n)}$ can be computed in terms of quantities determined
at lower orders. The first few orders are

\begin{equation} \mbox{diag}\left( [P^n,X^{(1)}]\right) = 0 \hspace{82mm} \end{equation}

\begin{equation} \mbox{diag}\left( [P^n,X^{(2)}] + [P^n,X^{(1)}]X^{(1)}
\right) = 0 \hspace{54mm} \label{second} \end{equation}

\begin{equation} \mbox{diag}\left( [P^n,X^{(3)}] + [P^n,X^{(2)}]X^{(1)} +
[P^n,X^{(1)}](X^{(2)} + X^{(1)}X^{(1)}) \right) = 0 \ . \end{equation}

\noindent At each order $s$ this is a linear inhomogeneous equation in the unknown
$X^{(s)}$. We record that

\begin{equation} B^{(s,n)} = \sum_{p=1}^{s-1}
\sum_{\substack{s_i\geq 1 \\ s_0 + \dots + s_p = s}}
[P^n,X^{(s_0)}]X^{(s_1)} \dots X^{(s_p)} \ . \label{Bsn} \end{equation}

\noindent We will need this result in the proofs of Theorems 4 and 5 below.

We end this section with two comments. The first concerns transposition of
Hadamard matrices: If $H$ is a
Hadamard matrix so is its transpose $H^{\rm T}$, a fact that will play a
role when we discuss explicit solutions later on. It is easy to show that

\begin{equation} H \rightarrow H^{\rm T} \hspace{5mm} \Leftrightarrow
\hspace{5mm} X \rightarrow FX^{\rm T}F^\dagger \ . \label{transpose}
\end{equation}

\noindent Hence transposition is slightly obscured by the discrete Fourier
transform.

The second comment concerns dephasing: In the introduction we observed that any
Hadamard matrix admits $2N-1$ free parameters that are in a sense trivial, and
that these trivial parameters can be removed by insisting that the first
row and the first column of the matrix has entries equal to $1/\sqrt{N}$
only. If we dephase the first row of the matrix $H$ we will obtain a
matrix $X$ whose first row contains zero entries only.
If we also dephase the first column of $H$ we find that $X$ takes the form

\begin{equation} X = \left[ \begin{array}{cccc} 0 & 0 & \dots & 0 \\
0 & \ & \ & \ \\ \vdots & \ & \tilde{X} & \\ 0 & \ & \ & \ \end{array}
\right] \ , \hspace{12mm} \sum_{j=1}^{N-1}\tilde{X}_{ij} = 0 \ .
\label{dephasing} \end{equation}

\noindent In our calculations we did not impose this condition, but we will
state our results in terms of the dimension of the set of dephased matrices.

\vspace{10mm}

{\bf 4. The homogeneous system}

\vspace{5mm}

\noindent We begin by solving the linear first order system ($n \neq 0$ mod
$N$)

\begin{equation} \mbox{diag}[P^n, X^{(1)}] = 0 \hspace{5mm} \Leftrightarrow
\hspace{5mm} X^{(1)}_{i+n,i} - X^{(1)}_{i,i-n} = 0 \ . \end{equation}

\noindent We observe that there is no condition on the diagonal elements of $X$,
and that the equalities that do arise connect only elements on some given displaced
diagonal. On the $n$th displaced diagonal we obtain the string of equalities

\begin{equation} X_{i,i-n} = X_{i+n,i} = X_{i+2n,i+n} = ...
\ . \end{equation}

\noindent It pays to think of the matrix as built up from displaced diagonals
rather than columns, so we introduce the parametrisation

\begin{equation} X_{i,j} = x_{i,j-i} \ . \end{equation}

\noindent Recall that all matrix indices obey modulo $N$ arithmetic.
Our string of equalities becomes

\begin{equation} x_{i,-n} = x_{i+n,-n} = x_{i+2n,-n} = ... \ . \end{equation}

\noindent Now let the greatest common divisor of $n$ and $N$ be denoted by $g$,
so that

\begin{equation} N = gr \ , \hspace{8mm} n = gs \ . \end{equation}

\noindent Taking the modulo $N$ arithmetic into account we see that the string of
equalities ends with an identity after $r$ steps,

\begin{equation} x_{i,-n} = ... = x_{i+rn,-n} \equiv x_{i+sN,-n} \equiv x_{i,-n} \ .
\label{string1} \end{equation}

\noindent There are $r-1$ non-trivial equalities here. Written in a more transparent fashion they are

\begin{equation} x_{i,-n} = x_{i+g,-n} = x_{i+2g,-n} = ... = x_{i+(r-1)g,-n}
\ . \label{string2} \end{equation}

\begin{figure}
        \centerline{ \hbox{
                \epsfig{figure=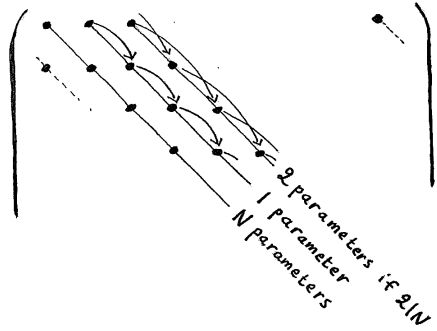,width=65mm}}}
        \caption{\small The proof of theorem 2; one should think of
        a matrix in terms of its diagonals, not in terms of its rows
        or columns.}
        \label{fig:defect1b}
\end{figure}

\noindent See Fig. \ref{fig:defect1b}.
This means that the $n$th displaced diagonal consists of $g$
sets of $r$ identical elements. Recalling that the elements on the main
diagonal are unrestricted it follows that the number of free parameters
at first order is

\begin{equation} \sum_{n=0}^{N-1} \mbox{gcd}(n,N) =
N + \sum_{n=1}^{N-1}\mbox{gcd}(n,N) \ . \end{equation}

\noindent This agrees with the result (\ref{TaZy}) due to Tadej
and \.Zyczkowski \cite{ZT}, except that at this stage our parameters are
complex since we have not yet imposed condition (\ref{ETT}). To linear
order condition (\ref{ETT}) evidently means that we end up with exactly
this number of real parameters, so the agreement is in fact complete.
If $N = p$ is a prime number all elements on each displaced diagonal are set equal.
In this case then there are exactly $2N-1$ free parameters in the solution, $N$
of them coming from the unrestricted main diagonal, which means that the dephased
Fourier matrix is isolated, not belonging to a continuous family of dephased Hadamard
matrices.

In effect we have proved

\

\noindent \underline{Theorem 2}: {\sl At any order, the homogeneous solution to
our linear system can be written as}

\begin{equation} X^{(s)}_{i,j} = x^{(s)}_{i \ {\rm mod \ gcd}(j-i,N),j-i
\ {\rm mod} \ N} \ . \label{losning} \end{equation}

\

\noindent An illustration of the proof is given in Fig. 1. If $N = p$ is a
prime number $X$ is a circulant matrix, except that its main diagonal is
unrestricted. Its dephased form is zero. A more interesting example is that
of $N = 6$, for which the first order solution is

\begin{equation} X^{(1)} = \left( \begin{array}{cccccc} x_{0,0} & x_{0,1} & x_{0,2} &
x_{0,3} & x_{0,4} & x_{0,5} \\ x_{0,5} & x_{1,0} & x_{0,1} & x_{1,2} & x_{1,3}
& x_{1,4} \\ x_{0,4} & x_{0,5} & x_{2,0} & x_{0,1} & x_{0,2} & x_{2,3} \\
x_{0,3} & x_{1,4} & x_{0,5} & x_{3,0} & x_{0,1} & x_{1,2} \\
x_{0,2} & x_{1,3} & x_{0,4} & x_{0,5} & x_{4,0} & x_{0,1} \\
x_{0,1} & x_{1,2} & x_{2,3} & x_{1,4} & x_{0,5} & x_{5,0} \end{array} \right) \ .
\label{6=N} \end{equation}

\noindent When we remove the trivial parameters by dephasing the Hadamard
matrix this is

\begin{equation} X^{(1)} =
\left( \begin{array}{cccccc} 0 & 0 & 0 & 0 & 0 & 0 \\ 0 & d_1 & 0 & x_{1,2} &
x_{1,3} & x_{1,4} \\ 0 & 0 & d_2 & 0 & 0 & x_{2,3} \\
0 & x_{1,4} & 0 & d_3 & 0 & x_{1,2} \\
0 & x_{1,3} & 0 & 0 & d_4 & 0 \\
0 & x_{1,2} & x_{2,3} & x_{1,4} & 0 & d_5 \end{array} \right) \ .
\label{N=6} \end{equation}

\noindent Now the diagonal elements $d_i$ are dependent parameters since
dephasing the Hadamard matrix means that the column sums of $X$ vanish, so the
number of free parameters equals
$4$. The affine Fourier family (see Appendix A) is obtained by setting
$x_{1,3} = x_{2,3} = 0$, and its transpose by setting $x_{1,2} = x_{1,4} = 0$; to see
what eq. (\ref{transpose}) implies for $X$ when $H$ is transposed it is
helpful to begin with the observation that circulant matrices are diagonalised
by the Fourier matrix.

\vspace{10mm}

{\bf 5. The heterogeneous systems}

\vspace{5mm}

\noindent The question now is whether there
are non-trivial consistency conditions on the linear systems that appear when
we insert the expansion (\ref{expansion}) in eq. (\ref{TVA}).
If so the first order result is misleading, and the true dimension of the
solution space is lower than the linear result suggests it should be. If
$N$ is a prime or a power of a prime we know---see section 1---that this
cannot happen, but for all other choices of $N$ we move on unknown ground.

The equations that we are faced with take the form (\ref{higher}). Our aim
is to bring their solution to a form precise enough to enable us to
implement it in Mathematica. This is achieved in the following two theorems.

\

\noindent \underline{Theorem 3}: {\sl The general solution to the heterogeneous
linear system (\ref{higher}) is}

\begin{equation} X^{(s)}_{i,j} = x^{(s)}_{i \ {\rm mod} \ {\rm gcd}(j-i,N),j-i}
+ \sum_{q=0}^{m[i,j,N]-1}B^{(s,i-j)}_{i+q(i-j),i+q(i-j)} \ ,
\label{solution} \end{equation}

\noindent {\sl subject to the consistency conditions}

\begin{equation} \sum_{q=0}^{N/{\rm gcd}(n,N) - 1}B^{(s,n)}_{i +
q \ {\rm gcd}(n,N),i+q \ {\rm gcd}(n,N)} = 0 \ ,
\label{cons} \end{equation}

\noindent {\sl where $n \in \{ 1, \dots , N-1 \}$,  $i \in \{ 0, \dots ,
\mbox{gcd}(n,N) - 1\}$ in the consistency conditions, and the $x_{i,j-i}$ are
free parameters. The integer $m = m[i,j,N]$ is given by}

\begin{equation} m = \frac{i \ \mbox{mod gcd}(i-j,N)-i}{\mbox{gcd}(i-j,N)}
\left( \frac{i-j}{\mbox{gcd}(i-j,N)}\right)^{-1} \ \mbox{mod} \
\frac{N}{\mbox{gcd}(i-j,N)}  \ , \label{modinverse} \end{equation}

\noindent {\sl where the inverse is the multiplicative inverse modulo
$N/\mbox{gcd}(i-j,N)$.}

\

\noindent In words, the consistency conditions at order $s$ say that for each
displaced diagonal labelled by $n$ the $B^{(s,n)}$ sum to zero when summed over
a set of positions with equal values of the parameters in the homogeneous
solution.

\

\noindent \underline{Proof}: Using explicit matrix indices, always taken
modulo $N$, eq. (\ref{higher}) becomes

\begin{equation} X^{(s)}_{i,i-n} = X^{(s)}_{i+n,i} + B^{(s,n)}_{ii} \ .
\end{equation}

\noindent Setting $n = i-j$ we find by iterating in $m$ steps that

\begin{equation} X^{(s)}_{i,j} = X^{(s)}_{i+m(i-j),j+m(i-j)} + \sum_{q=0}^{m-1}
B^{(s,i-j)}_{i+q(i-j),i+q(i-j)} \ . \label{alongdiag} \end{equation}

\noindent When

\begin{equation} m(i-j) = 0 \ \ \mbox{mod} \ N \hspace{5mm} \Leftrightarrow
\hspace{5mm} m = \frac{kN}{\mbox{gcd}(i-j,N)} \ , \ \ k \in {\bf Z} \end{equation}

\noindent we obtain a consistency condition which---after a slight reordering
of the sum---is exactly eq. (\ref{cons}). This is the same reordering that
was used to go from (\ref{string1}) to (\ref{string2}).

Now we must find a particular solution of the heterogeneous system. We begin
by choosing all matrix elements corresponding to independent parameters in the
homogeneous solution to zero, beginning with all elements in the upmost row and
continuing downwards until the independent elements are exhausted. For this
particular solution

\begin{equation} i \mbox{\ mod gcd}(i-j,N) = i \hspace{5mm}
\Rightarrow \hspace{5mm} X^{(s)}_{i,j} = 0
\ . \label{villkor} \end{equation}

\noindent Coming back to eq. (\ref{alongdiag}), which relates elements along
the $(i-j)$th diagonal in steps of gcd$(i-j,N)$, we see that we obtain a
particular solution fully determined by the lower order solution provided that
the first term on the right hand side vanishes. Given the conditions (\ref{villkor})
that we already imposed this will be ensured if we can
find (for each $i,j,N$) an integer $m$ such that

\begin{equation} i + m(i-j) = i + m(i-j) \
\mbox{mod gcd}(i-j,N)  = i \ \mbox{mod gcd}(i-j,N) \ . \end{equation}

\noindent This is equivalent to

\begin{equation} \left\{ m\frac{i-j}{\mbox{gcd}(i-j,N)} = \frac{i \ \mbox{mod
gcd}(i-j,N) - i}{\mbox{gcd}(i-j,N)}\right\} \ \mbox{mod} \frac{N}{\mbox{gcd}
(i-j,N)} \ . \end{equation}

\noindent The solution in the range $\{ 0, ... , N/\mbox{gcd}(i-j,N)\}$
is given in eq. (\ref{modinverse}). The modular
multiplicative inverse used there
exists since the numbers $(i-j)/\mbox{gcd}(i-j,N)$ and $N/\mbox{gcd}(i-j,N)$ are
coprime. Using this value of $m$ in eq. (\ref{alongdiag}) gives the particular
solution, and we arrive at the general solution (\ref{solution}) by adding the
general solution of the homogeneous equations. $\Box$

\

To illustrate the proof we give the particular solution for $N = 6$:

{\small

\begin{equation} \left( \begin{array}{cccccc} 0 & 0 & 0 & 0 & 0 & 0 \\ \\
\sum_{0}^4B_{1+q}^{(1)} & 0 & B_{1}^{(5)} & 0 & 0 & 0 \\ \\
\sum_{0}^1B_{2+2q}^{(2)} & \sum_{0}^3B_{2+q}^{(1)} & 0
& \sum_{0}^1B_{2-q}^{(5)} & B_{2}^{(4)} & 0 \\ \\
B_{3}^{(3)} & \sum_{0}^1B_{3+2q}^{(2)} & \sum_{0}^2B_{3+q}^{(1)}
& 0 & \sum_{0}^2B_{3-q}^{(5)} & B_{3}^{(4)} \\ \\
\sum_{0}^1B_{4-2q}^{(4)} & B_4^{(3)} & B_4^{(2)} & \sum_{0}^1B_{4+4q}^{(1)} &
0 & \sum_{0}^3B_{4-q}^{(5)} \\  \\ \sum_0^4B_{5-q}^{(5)} & \sum_0^1B_{5-2q}^{(4)}
& B_{5}^{(3)} & B_{5}^{(2)} & B_5^{(1)} & 0
\end{array} \right)  \end{equation}

}

\noindent where (for typographical reasons) we let $B^{(s,n)}_{i,i}$
be denoted by $B_i^{(n)}$.

To make use of theorem 3 we must compute the inhomogeneous term $B^{(s,n)}_{i,i}$.
This is achieved in the next theorem.

\

\noindent \underline{Theorem 4}: {\sl The inhomogeneous part $B^{(s,n)}$ is given
iteratively by}

\begin{equation} B^{(1,n)} = 0 \end{equation}

\begin{equation} B^{(s,n)} = \sum_{r=1}^{s-1}\left( [P^n,X^{(r)}] + B^{(r,n)}\right) X^{(s-r)} \ . \end{equation}

\

\noindent The proof is straightforward, using expression (\ref{Bsn}).

It is hard to continue the discussion with any generality, but we have been
able to prove that the consistency conditions arising at second
order, eqs. (\ref{second}), are identically satisfied for all $N$. We give a brief
sketch of a proof in Appendix C. At third order the
consistency conditions do break down for some values of $N$, as will be
discussed in section 7.

\vspace{10mm}

{\bf 6. The unitarity equations}

\vspace{5mm}

\noindent We must now impose the unitarity condition (\ref{ETT}) on the solutions
obtained in section 5. When $X$ is expanded in a power series the unitarity
condition at order $s$ reads

\begin{equation} X^{(s)} + \left( X^{(s)}\right)^\dagger = \sum_{r=1}^{s-1}X^{(s-r)}
\left( X^{(r)}\right)^\dagger \ . \label{Xdagger} \end{equation}

\noindent We insert the solution

\begin{equation} X^{(s)}_{i,j} = x_{i,j-i}^{(s)} + H^{(s)}_{i,j}(X^{(1)}, \dots ,
X^{(s-1)}) \ , \end{equation}

\noindent where $H_{i,j}^{(s)}$ is the heterogeneous part as given in theorem 3,
and $x^{(s)}_{i,j-i}$ is short for $x^{(s)}_{i \ {\rm mod \ gcd}(j-i,N),j-i
\ {\rm mod} \ N}$.
The equations to be solved are now

\begin{equation} x^{(s)}_{i,j-i} + \bar{x}^{(s)}_{i,i-j} = F_{i,j}^{(s)} \ ,
\end{equation}

\noindent where complex conjugation is denoted by an overbar. We used
$i \ {\rm mod \ gcd}(j-i,N) = j \ {\rm mod \ gcd}(j-i,N)$,
and we also defined

\begin{equation} F^{(s)} = \sum_{r=1}^{s-1}X^{(s-r)}
\left( X^{(r)}\right)^\dagger - H^{(s)} - \left( H^{(s)}\right)^\dagger
\ . \end{equation}

\noindent Note that this depends on the solution only to orders less than $s$.

\

\noindent \underline{Theorem 5}: {\sl The matrix $H = ({\bf 1}-X)F$ is unitary
order by order in the perturbative expansion provided that the following
conditions are imposed on its matrix elements:}

\begin{equation} \mbox{Re}\left[ x_{i,0}^{(s)}\right] = \frac{1}{2}F_{i,i}^{(s)}
\end{equation}

\begin{equation} \mbox{Re}\left[ x_{i \mbox{{\small \ mod}}\ N/2,N/2}^{(s)}\right] = \frac{1}{2}F_{i,i+N/2}^{(s)} \hspace{5mm} \mbox{\sl if} \ N \ \mbox{\sl is even}
\label{ditt} \end{equation}

\begin{equation} x^{(s)}_{i \mbox{{\small \ mod gcd}}(n,N),-n} =
F_{i,j}^{(s)} - \bar{x}^{(s)}_{i \mbox{{\small \ mod gcd}}(n,N),n} \hspace{5mm}
\mbox{\sl for} \ 0 < n < N/2 \ . \label{datt} \end{equation}

\

\noindent \underline{Proof}: Most of the calculation is already done, but in
eqs. (\ref{ditt}-\ref{datt}) some equations are repeated as $i$ runs through
its $N$ values and we must prove consistency, namely that $F^{(s)}$ solves the
same linear system as does the homogeneous solution. Thus we require

\begin{equation} \mbox{diag}[P^n,F^{(s)}] = 0 \ . \end{equation}

\noindent Using the fact that diag$\{ [P^n,H^{(s)}]\} = - \mbox{diag}\{ B^{(s,n)}\}$ 
and recalling that $(P^n)^\dagger = P^{-n}$,  a calculation shows
that

\begin{eqnarray} \mbox{diag}[P^n,F^{(s)}] = \mbox{diag}\left( \sum_{r=1}^{s-1}
[P^n,X^{(s-r)}]X^{(r)\dagger} + B^{(s,n)} \right) - \ \ \ \nonumber \\ \ \label{noll} \\
\hspace{35mm} - \mbox{diag}\left( \sum_{r=1}^{s-1}
[P^{-n},X^{(s-r)}]X^{(r)\dagger} + B^{(s,-n)} \right)^\dagger \ .  \nonumber \end{eqnarray}

\noindent Next we show directly from eq. (\ref{Xdagger}) that

\begin{equation} X^{(s)\dagger} = - \sum_{p=1}^s
\sum_{\substack{s_i\geq 1 \\ s_1 + \dots + s_p = s}}
X^{(s_1)}\dots X^{(s_p)} \ . \end{equation}

\noindent Using this result, and using eq. (\ref{Bsn}) to substitute for $B^{(s,n)}$,
we see that the right hand side of eq. (\ref{noll}) vanishes, as was to be shown. $\Box$

\

We illustrate this by imposing unitarity on the first order result for $N = 6$
given in eq. (\ref{6=N}). It now reads

\begin{equation} X^{(1)} = \left( \begin{array}{rrrrrr} y_{0,0} & x_{0,1} & x_{0,2} &
y_{0,3} & -\bar{x}_{0,2} & -\bar{x}_{0,1} \\ -\bar{x}_{0,1} & y_{1,0} & x_{0,1}
& x_{1,2} & y_{1,3} & -\bar{x}_{1,2} \\
-\bar{x}_{0,2} & -\bar{x}_{0,1} & y_{2,0} & x_{0,1} & x_{0,2} & y_{2,3} \\
y_{0,3} & -\bar{x}_{1,2} & -\bar{x}_{0,1} & y_{3,0} & x_{0,1} & x_{1,2} \\
x_{0,2} & y_{1,3} & -\bar{x}_{0,2} & -\bar{x}_{0,1} & y_{4,0} & x_{0,1} \\
x_{0,1} & x_{1,2} & y_{2,3} & -\bar{x}_{1,2} & -\bar{x}_{0,1} & y_{5,0}
\end{array} \right) \ . \label{unitary} \end{equation}

\noindent The matrix elements denoted $y_{i,j}$ are purely imaginary. The
others are complex and related in pairs by complex conjugation.

The important point is that the unitarity condition simply cuts the number of free
parameters in half; in other words if there were $d$ complex parameters in the
perturbative solution of eq. (\ref{TVA}) then we end up with $d$ real parameters
after imposing unitarity. The difficult consistency conditions appear only
in the solution of eq. (\ref{TVA}).

We are unable to show in general that unitarity also cuts
the number of variables fixed by a non-trivial consistency condition in half.
Still, in section 8 we prove that this indeed happens for the particular case
of $N = 12$. We believe that the issue must be dealt with on a case-by-case basis.

\vspace{10mm}

{\bf 7. When do the consistency conditions break down?}

\vspace{5mm}

\noindent As was mentioned at the end of section 5 the consistency conditions
always hold to second order. For third order and higher we have used Mathematica
to perform the calculation numerically, using random values for the free
parameters. For some dimensions to be discussed below the calculation was also
performed symbolically. When $N$ is a power of a prime we found no breakdown in
the consistency conditions to the orders we checked. This had to be so because
in this case there do exist affine families of Hadamard matrices with their
dimension given by the linear defect $d_1$. For $N = 2\cdot 3 = 6$ we
computed (numerically) the higher order contributions to order 100 in the
perturbation series, without encountering any breakdown of the consistency
conditions. There seemed to be no point in going further. The calculation
clearly supports the extant conjecture that the $N = 6$ Fourier matrix belongs
to a smooth 4-parameter family of dephased Hadamard matrices.

For all other choices of $N \leq 100$ we found that the consistency conditions
do break down at some order $s$, according to a definite pattern. Let $p_1,
p_2, p_3, \dots$ be different prime numbers. Then the consistency conditions

\begin{itemize}

\item{hold to all orders if $N$ is a power of a prime (and to order 100 if $N = 6)$}

\item{break at order 11 if $N = 10$}

\item{break at order 7 if $N= 2p_1$ and $p_1 > 5$ is odd}

\item{break at order 5 if $N = p_1p_2$ is odd}

\item{break at order 4 if $N = p_1^{k_1}p_2^{k_2}$ and $k_1k_2 > 1$}

\item{break at order 3 if $N = p_1^{k_1}p_2^{k_2}p_3^{k_3}$.}

\end{itemize}

\noindent For $N \leq 100$ there are no examples of an integer that is a
product of four different primes, but we did check that the consistency conditions
break at order 3 for $N = 2\cdot 3\cdot 5 \cdot 7 = 210$. Again, Appendix C
contains a proof that the consistency conditions always hold to second
order.

It is very encouraging that an orderly
pattern emerges.

\vspace{10mm}

{\bf 8. The $N = 12$ case}

\vspace{5mm}

\noindent What we really want to know is the dimension of the solution space
to all orders. We will discuss the case $N = 12$ in detail. The defect at linear
order is $d_1 =17$, and it is also known that the Fourier matrix belongs to
seven distinct affine families of dimension $d_A = 9$ \cite{TZ}. Therefore
we know at the outset that the dimension of the dephased solution space lies
between these bounds.

When the calculation is continued beyond linear order we find that
the consistency conditions for the linear system break down at fourth order in
the perturbation, resulting in a set of fourth order polynomials to be solved.

There are 40 variables in the first order solution. 23 of those are trivial
(since they determine the $2N-1$ free phases) so we expect the consistency
conditions to depend on 17 variables. There are only 13 conditions that do
not vanish automatically, and close inspection shows that only 13 linear
combinations of the variables enter these equations. To be precise, the
consistency conditions are fourth order polynomials in the 13 variables

\begin{equation} \begin{array}{ll} x_2 = x_{0,2}-x_{1,2} \\ x_{10} = x_{0,10} - x_{1,10}
\\ x_{4a} = x_{0,4} - x_{2,4} \\ x_{4b} = x_{1,4} - x_{3,4}
\\ x_{8a} = x_{0,8} - x_{2,8} \\ x_{8b} = x_{1,8} - x_{3,8}
\\ x_{6a} = x_{0,6} - x_{3,6} \\ x_{6b} = x_{4,6} - x_{1,6}
\\ x_{6c} = x_{2,6} - x_{5,6}  \end{array} \hspace{10mm}
\begin{array}{ll} x_{3a} = 2x_{0,3} - x_{1,3} - x_{2,3} \\ x_{3b} =
2x_{1,3} - x_{0,3} - x_{2,3} \\ x_{3c} = 2x_{2,3} - x_{0,3} - x_{1,3}
\\ x_{9a} = 2x_{0,9} - x_{1,9} - x_{2,9} \\ x_{9b} =
2x_{1,9} - x_{0,9} - x_{2,9} \\ x_{9c} = 2x_{2,9} - x_{0,9} - x_{1,9} & \ .
\end{array} \label{variables} \end{equation}

\noindent Note that there are only 13 independent variables, because
$x_{3a} + x_{3b} + x_{3c} = x_{9a} + x_{9b} + x_{9c} = 0$.

We next define the auxiliary polynomials

\begin{equation} p_1 = x_{3a}^2 + x_{9a}^2 \hspace{8mm} p_2 = x_{3b}^2 + x_{9b}^2
\hspace{8mm} p_3 = x_{3c}^2 + x_{9c}^2 \end{equation}

\begin{equation} p_4 = x_{4a}^2 - x_{4b}^2 \hspace{6mm} p_5 = x_{8a}^2 - x_{8b}^2
\hspace{6mm} p_6 = x_{4a}x_{8a} - x_{4b}x_{8b} \ . \end{equation}

\noindent The complete set of consistency conditions now takes the form

\begin{equation} p_1p_6 = p_2p_6 = p_3p_6 = 0 \end{equation}

\begin{equation} (p_1+p_2+p_3)p_4 = (p_1+p_2+p_3)p_5 = 0 \end{equation}

\begin{equation} x_{4a}x_{10}(p_1+p_2+p_3) + x_{8a}(x_{6a}p_1 + x_{6b}p_2 +
x_{6c}p_3) = 0 \end{equation}

\begin{equation} x_{4b}x_{10}(p_1+p_2+p_3) + x_{8b}(x_{6a}p_1 + x_{6b}p_2 +
x_{6c}p_3) = 0 \end{equation}

\begin{equation} x_{8a}x_{2}(p_1+p_2+p_3) + x_{4a}(x_{6a}p_1 + x_{6b}p_2 +
x_{6c}p_3) = 0 \end{equation}

\begin{equation} x_{8b}x_{2}(p_1+p_2+p_3) + x_{4b}(x_{6a}p_1 + x_{6b}p_2 +
x_{6c}p_3) = 0 \end{equation}

\begin{equation} 3x_{3a}(x_{10}p_4+x_2p_5) + 2p_6\left( x_{3a}(2x_{6a} + x_{6c}) + x_{3b}
(x_{6c} - x_{6b})\right) = 0 \ \ \end{equation}

\begin{equation} 3x_{3b}(x_{10}p_4+x_2p_5) + 2p_6\left( x_{3b}(2x_{6b} + x_{6c}) + x_{3a}
(x_{6c} - x_{6a})\right) = 0 \ \ \end{equation}

\begin{equation} 3x_{9a}(x_{10}p_4+x_2p_5) + 2p_6\left( x_{9a}(2x_{6a} + x_{6c}) + x_{9b}
(x_{6c} - x_{6b})\right) = 0 \ \ \end{equation}

\begin{equation} 3x_{9b}(x_{10}p_4+x_2p_5) + 2p_6\left( x_{9b}(2x_{6b} + x_{6c}) + x_{9a}
(x_{6c} - x_{6a})\right) = 0 \ . \end{equation}

\noindent There are three ways of satisfying these equations.

Solutions of type 1 are obtained if $p_4 = p_5 = p_6 =0$. This implies that

\begin{equation} \begin{array}{l} x_{4a} = x_{4b} \\ x_{8a} = x_{8b} \end{array}
\hspace{7mm} \mbox{or} \hspace{7mm} \begin{array}{ll} x_{4a} = - x_{4b} \\ x_{8a} = - x_{8b} & \ . \end{array} \end{equation}

\noindent Two polynomial equations remain,
so the dimension drops by 4. Note that the four conditions

\begin{equation} x_{4a} = x_{4b} = x_{8a} = x_{8b} = 0 \label{typI} \end{equation}

\noindent solves the entire system. Call this special case the type I
solution.

Solutions of type 2 are obtained if $p_1 = p_2 = p_3 = 0$. This implies that

\begin{equation} \begin{array}{l} x_{3a} = ix_{9a} \\ x_{3b} = ix_{9b} \end{array}
\hspace{7mm} \mbox{or} \hspace{7mm} \begin{array}{ll} x_{3a} = - ix_{9a} \\ x_{3b} =
- ix_{9b} & \ . \end{array} \end{equation}

\noindent Two polynomial equations remain,
so the dimension again drops by 4. Note that the four conditions

\begin{equation} x_{3a} = x_{3b} = x_{9a} = x_{9b} = 0 \label{typII} \end{equation}

\noindent solves the entire system. Call this special case the type II
solution.

When neither $p_1,p_2,p_3$ nor $p_4,p_5$ are all zero
the equations imply

\begin{eqnarray} \left\{ \begin{array}{ll} p_1+p_2+p_3 = 0 \\ \\ p_6 = 0 \end{array}
\right. \hspace{21mm} \label{L} \\
\left\{ \begin{array}{ll} x_{6a}p_1 + x_{6b}p_2 + x_{6c}p_3 = 0 \\ \\
x_{10}p_4 + x_2p_5 = 0 & \ . \end{array} \right. \label{R}
\end{eqnarray}

\noindent Hence the consistency conditions impose 4 independent conditions on our variables, meaning that the defect again drops from 17 to 13. Call these solutions type 3.

Solutions of type 2 are obtained by transposition from solutions of type 1,
and similarly for types II and I. Some work
is needed to verify this since transposing the Hadamard matrix leads to a
slightly unobvious operation on the matrix $X$; see (\ref{transpose}).
Solutions of type 3 on the other hand form
a self-cognate family (at least when considered as an algebraic variety),
since transposition interchanges eqs. (\ref{L}) and eqs. (\ref{R}).

Let us now impose unitarity, following section 6. The consistency
conditions we have encountered are quartic polynomials in the first order
matrix elements, and we begin by imposing unitarity to first order. Then
$x_{i,6}$ become purely imaginary, and the remaining
parameters become related by complex conjugation ($x_{0,2} + \bar{x}_{0,10} = 0$,
\dots , $x_{0,3} + \bar{x}_{0,9} = 0$, \dots , $x_{0,4} + \bar{x}_{0,8} = 0$, \dots ).
Compare the simpler example given in eq. (\ref{unitary}).
This means that $p_1,p_2,p_3,p_6$ are real, while $p_5 = \bar{p}_4$.
There are 8 polynomial equations remaining. Solving them one ends up with
the same solutions as one obtains by imposing unitarity on the solutions of
the complex equations. Thus the possible difficulty mentioned at the end of
section 6 does not arise in this case.

Having solved the consistency conditions that arise at fourth order, we must
now address the question whether the 13-dimensional dephased solutions found
at fourth order are truly 13-dimensional, or whether their
dimensions drop due to consistency conditions appearing in higher orders.
To this end we continued the calculation numerically to order 11 for the solutions of
type I and II. No further breakdown of any consistency condition was observed.
For solutions of type 1 on the other hand a further breakdown of the consistency
conditions occurs at order 6, which means that the dimension of this family of
solutions drops below 13. Type 2 must behave similarly. For type 3 there is a
further breakdown of the consistency conditions already at order 5. Thus the
conclusion from these calculations is that the solutions of types 1, 2, and 3
have dimension smaller than 13. They may be entirely spurious. For types I
and II no such conclusion can be drawn, the calculation simply shows that
their dimension is at most 13.

Fortunately we can bound the dimension of the solutions from below as well.
As noted by Karlsson \cite{Karlsson2} the Di\c{t}\u{a} construction \cite{Dita}
(see Appendix A) allows us to construct a 13-dimensional family of $N = 12$
dephased Hadamard matrices once a 4-dimensional family of $N = 6$ dephased
Hadamard matrices is known. Indeed,
for any Hadamard matrices $H_1$ and $H_2$ and any diagonal unitary matrix $D$ of
order $N$ the matrix

\begin{equation} H = \left( \begin{array}{cr} H_1 & DH_2 \\ H_1 & - DH_2 \end{array}
\right) \label{Ditaigen} \end{equation}

\noindent is a Hadamard matrix of order $2N$. If $H_1$ and $H_2$ include the
Fourier matrix of order $N$ this family includes a matrix which is permutation
equivalent to the Fourier matrix of order $2N$. Given that the diagonal unitary
$D$ contributes 5 free phases to the dephased matrix $H$, and that the existence
of a 4 dimensional family for $N = 6$ ensures that $H_1$ and $H_2$ can contribute
4 phases each, this implies that there exists at least one 13-dimensional
family for $N = 12$. The only candidates for such a family are the type I and II
solutions. (Since they are related by transposition the existence of one implies
the existence of the other.) Therefore it is only the slight uncertainty concerning
$N = 6$ that prevents us from stating as a theorem that these 13-dimensional
solutions must exist to all orders.

This conclusion receives strong support from a different direction. For $N = 12$
it is known that there exist exactly 7 affine families of dimension
9 \cite{TZ}. Our perturbative solutions contain these families. The two
13-dimensional solutions of types I and II intersect in a 9-dimensional
affine family called $F_{12A}$ \cite{TZ}, and each type contains one representative
of each of three pairs of affine families related by transposition.
For the self-cognate family $F_{12A}$ the parameters introduced by Tadej and
\.Zyczkowski \cite{TZ} are explicitly given by

\begin{equation} \begin{array}{ll} a = x_{1,2}(\omega^5 + \omega^9) + x_{1,4}(\omega^7 + \omega^9) +
c.c. + 2\alpha_1 \\
b = x_{1,2}(\omega^7 + \omega^9) + x_{1,4}(\omega^9 + \omega^{11}) +
c.c.  \\ c = 2x_{1,2}\omega^9 + c.c. + 2\alpha_1 \\
d = x_{1,2}(\omega^9 + \omega^{11}) + x_{1,4}(\omega^7 + \omega^9) +
c.c.  \\ e = x_{1,2}(\omega + \omega^9) + x_{1,4}(\omega^9 + \omega^{11}) +
c.c. + 2\alpha_1 \\ f = 2\alpha_2 \\
g = x_{1,2}(\omega^5 + \omega^9) + x_{1,4}(\omega^7 + \omega^9) + c.c.
+ 2\alpha_3 \\ h = 2\alpha_4 \\ i = x_{1,2}(\omega^5 + \omega^9) + x_{1,4}(\omega^7 + \omega^9) + c.c.
+ 2\alpha_5 & \ , \end{array} \end{equation}

\noindent where $\alpha_i = -ix_{i,6}$ is real and $c.c.$ denotes the complex
conjugate of the preceding terms. We do not give the choices that reproduce the
remaining affine families here.
The picture that emerges is that of Fig. 2.

\begin{figure}
        \centerline{ \hbox{
                \epsfig{figure=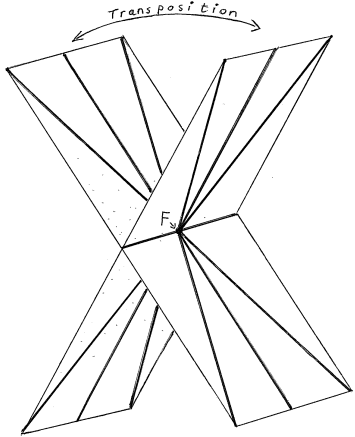,width=50mm}}}
        \caption{\small The type I and II solutions for $N = 12$ form
        two sheets related by transposition. The sheets intersect in
        an affine family, and each sheet contains three additional
        affine families all passing through the Fourier matrix, where the linear
        span of the tangent vectors of the two sheets has a dimension equal
        to its linear defect.}
                \label{fig:defect2}
\end{figure}

Now we have computed the linear defect $d_1$ for a few randomly chosen members
of each affine family. For the self-cognate family we find $d_1 = 17$. We observe 
that $13 + 13 - 9 = 17$, which means that the dimension of the linear span of
the tangent vectors of the intersecting solutions of type I and II is equal
to the linear defect $d_1=17$ of the self-cognate family in which they intersect.
For the remaining affine families we find by sampling a large number of
members that $d_1 = 13$ (except at the Fourier matrix itself),
which is consistent with the fact that these families sit within a
13-dimensional family. It also provides some evidence that the solutions of
types I and II exist also far away from the Fourier matrix.

\vspace{10mm}

{\bf 9. $N = p_1p_2^2$, and other choices of $N$}

\vspace{5mm}

\noindent Let us assume that $N = p_1p_2^2$, where $p_1$ and $p_2$ are
prime numbers. We are then able to formulate a precise conjecture concerning
how the type I and II solutions found for $N = 12$ generalise. Thus we
suggest that when $N = p_1p_2^2$ there are always two types of solutions,
giving rise to families of Hadamard matrices related to each other by
transposition. The conjectured solutions are given by

\begin{equation} \mbox{Type I :} \hspace{9mm} x_{i,j} = x_{i+p_2,j} \hspace{5mm}
\mbox{if} \hspace{5mm}  \mbox{gcd}(j,N) = p_2^2 \ \label{Ia} \end{equation}

\begin{equation} \mbox{Type II :} \hspace{8mm} x_{i,j} = x_{i+1,j} \hspace{5mm}
\mbox{if} \hspace{5mm} \mbox{gcd}(j,N) = p_1 \ . \label{IIa} \end{equation}

\noindent The evidence for this conjecture is primarily that these conditions
solve all consistency conditions up to fourth order not only for
$N = 12$---where they are equivalent to eqs. (\ref{typI}) and (\ref{typII}),
respectively---but for all such $N \leq 50$ (namely for $N = 3\cdot 2^2,
5\cdot 2^2, 7\cdot 2^2, 11\cdot 2^2, 2\cdot 3^2, 5\cdot 3^2$, and $2\cdot 5^2$). 
But we can add some additional evidence.

First of all, although we have continued the calculation to higher orders
only for the case $N = 12$, we note that the existence of a four parameter
family for $N = 6$ implies the existence of 22-dimensional family for
$N = 18$, via the Di\c{t}\u{a} construction. Compare eq. (\ref{Ditaigen}). 
This is precisely the dimension implied by our conjecture in this case.

Moreover the overall picture gained in the $N = 12$ case seems to repeat itself.
In general, eqs. (\ref{Ia}) or (\ref{IIa}) impose $(p_1-1)p_2(p_2-1)$
conditions on the first order solution. If the conjecture is true it implies
that the type I and II solutions have dimension

\begin{equation} d = d_1 - N\left( 1 - \frac{1}{p_1}\right) \left( 1 -
\frac{1}{p_2}\right) \ , \end{equation}

\noindent where $d_1$ is the linear defect of the Fourier matrix. Using the
expression for $d_1$ that we quote in eq. (\ref{Slom}), the dimension $d$ can
alternatively be expressed as

\begin{equation} d = d_A + N\left( 1 - \frac{1}{p_1}\right) \left( 1 -
\frac{1}{p_2}\right) \ , \label{dimension?} \end{equation}

\noindent where $d_A$ is the dimension of the largest affine family
obtainable from the Di\c{t}\u{a} construction in these dimensions. See eq.
(\ref{dA}) in Appendix A for this. It follows that

\begin{equation} d = \frac{d_1+d_A}{2} \hspace{5mm} \Leftrightarrow \hspace{5mm}
d_1 = 2d - d_A \ . \end{equation}

\noindent What this equation says is that the conjectured dimension $d$ is just
right for the two solutions to intersect in a
self-cognate affine family of dimension $d_A$, in such a way that the linear
span of their tangent vectors equals the linear defect at their intersection.

Using the Di\c{t}\u{a} construction we obtained one self-cognate family and
three pairs of affine families related by transposition, all of dimension $d_A$,
for all $N = p_1p_2^2$. The details
are in Appendix A. For $N = 12$ it is known that these are all affine
families of maximal dimension stemming from the Fourier matrix \cite{TZ}.
For our seven examples with $N = p_1p_2^2 \leq 50$ we
checked that the self-cognate family sits in the intersection of the two
solutions of type I and II, while the other affine families are contained
in a single such solution. We also computed the defect along all these affine
families for some randomly chosen matrices, and found that their linear
defect equals $2d - d_A$ for the self-cognate family, and it equals
$d$ for the other families (except at the Fourier matrix itself). This
clearly supports the conjecture that the
solutions of type I and II survive to higher orders.

Finally, a word about the equations that we actually encounter.
As in the case of $N = 12$ we know that the fourth order consistency conditions
admit other solutions besides the ones that we denote types I and II. For
$N = 20$ we found the general solution at fourth order. In this case the equations
depend on the variables only through combinations similar to those given in eqs.
(\ref{variables}), on

\begin{equation} x_{0,5} - x_{1,5} - x_{3,5} + x_{4,5} \ , \hspace{6mm}
4x_{0,5} - x_{1,5} - x_{2,5} - x_{3,5} - x_{4,5} \ , \end{equation}

\noindent and on cyclic permutations of these. Besides the two 25 dimensional
solutions of types I and II there exists a 29-dimensional
solution which is self-cognate (as an algebraic variety). However, it may
well be that such additional solutions are subject to further restrictions at
higher orders, as is indeed the case for $N = 12$. They may be entirely
spurious.

About dimensions $N$ not of the form $p_1p_2^2$ (or $p^k$) we have little to say.
We have found some hints that eq. (\ref{dimension?}) may give the dimension
of a family of Hadamard matrices in somewhat greater generality,
notably if $N = p_1p_2^k$
and $k > 1$. The case $N = p_1p_2$ works in a different way however.

\vspace{10mm}

{\bf 10. Conclusions and speculations}

\vspace{5mm}

\noindent We have investigated the dimension of any smooth set of
$N\times N$ complex Hadamard matrices including the Fourier matrix.
Our method uses a perturbative expansion in which a linear system is
solved at each order. At each order a set of non-linear consistency
conditions on the lower order solution must be satisfied. An interesting
feature is that we first solve the complexified equations, and then
impose unitarity order by order on the solution.

We state our results for the set of dephased Hadamard matrices. If $N$
is a prime number this set consists of a single matrix, namely
the Fourier matrix itself. If $N$ is a power of a prime number the
dimension of the set is known from previous work, because there are explicit
constructions of affine families that saturate the bound given by the
first order calculation \cite{Dita}; see Appendix B for an explicit
construction.
For $N = 6$ our calculations support, to 100th order in perturbation
theory, the conjecture that the dimension equals 4. If this is so,
it follows that the dimension for $N = 12$ is at least 13. Our
calculations---carried to 11th order in perturbation theory---support
this number, and indeed they prove that the dimension cannot exceed 13
due to consistency conditions that appear at order 4.
Moreover an appealing structure emerged, in which the known affine families
find a natural place. First order calculations along these families
provide evidence that the 13-dimensional families remain 13-dimensional
also far away from the Fourier matrix.

For all non-prime power values of $N$ with the single exception
of $N = 6$, we found that non-trivial consistency conditions arise in the
calculation. Our conclusion is that the consistency conditions on the
first order solution are identically satisfied if $N = 6$ or if $N$ is
a prime power, that non-trivial consistency conditions appear at order
11 if $N = 10$, at order 7 if $N$ is twice an odd prime with $N > 10$,
at order 5 if $N$ is a product of two odd primes, at order 4 if $N$ is a
product of two primes with at least one prime factor repeated, and at
order 3 if $N$ contains three different prime factors. Our evidence for
this statement was described in section 7. Based on it we confidently
suggest that a systematic understanding of these calculations is possible
for arbitrary $N$, even though this is out of our reach at the moment.

It is interesting to observe that 6, and to some extent 10, dimensions stand
out as being very special. The conjecture that complete sets of Mutually
Unbiased Bases do not exist in any non-prime power dimension rests
on numerical searches for $N = 6$; see ref. \cite{Raynal} and references
therein.

When the consistency conditions break down we know that the dimension of the
solution space is less than that suggested by the linear defect. However,
in order to see by how much the dimension drops it is necessary to solve these
conditions---which are multivariate polynomial equations. We were able to
deal with this problem for altogether seven examples where $N$ is of the
form $p_1p_2^2$.
Based on the results described in sections 8 and 9 we were then able to
conjecture the form of a solution for all $N= p_1p_2^2$. The conjectured
solution consists of two families of dimension

\begin{equation} d = \frac{d_1+d_A}{2} = 3N - 3p_1p_2 - 2p_2^2 + p_2 + 1 \ ,
\end{equation}

\noindent where $d_1$ is the linear defect and $d_A$ the maximal dimension of
known affine families. These two families are related by transposition and
intersect in a self-cognate affine family of dimension $d_A$, as illustrated
in Fig. 2. There are other affine families contained within a
single branch of the solution.

We are not sure whether other solutions exist, or not. Indeed this is
as far as we have been able to go. We have no hints for what a
completely general formula for the dimension of the set of Hadamard
matrices containing the Fourier matrix will look like. Nevertheless
the evidence strongly suggests that one can be found.
We find it intriguing that our results depend on the number
theoretical properties of the dimension in such an intricate way.


\

\

\noindent \underline{Acknowledgements}: We thank Markus Grassl,
Bengt Karlsson, {\L}ukasz Skow- ronek, Feri Sz{\"o}ll{\H{o}}si, Wojtek
Tadej, and Karol \.Zyczkowski for sharing their knowledge, and a not 
so anonymous referee for exceptionally detailed criticism. IB is
supported by the Swedish Research Council under contract VR 621-2010-4060.

\vspace{10mm}

{\bf Appendix A: Background information}

\vspace{5mm}

\noindent {\it Equivalences}: In the classification of complex Hadamard matrices
two matrices $H_1,H_2$ are regarded as equivalent, written $H_1 \approx H_2$, if
there exist diagonal matrices $D_L,D_R$ and permutation matrices $P_L,P_R$ such that

\begin{equation} H_2 = D_LP_LH_1P_RD_R \ . \end{equation}

\noindent This is a natural equivalence relation in many applications \cite{Haagerup}.
If the Hadamard matrices are given in dephased form only discrete equivalences
remain. In practice it is hard to take all of them into account.

For composite $N = N_1N_2$ one finds that the Fourier matrix
$F_N \approx F_{N_1}\otimes
F_{N_2}$ if and only if $N_1$ and $N_2$ are relatively prime. Since $F_N$ is
the character table of the cyclic group $Z_N$ this follows from a well
known fact about cyclic groups.

\

\noindent {\it The linear defect}: Here we just want to mention that the
formula for the linear defect, eq. (\ref{defect}), can be rewritten in
the form

\begin{equation} d_1 = \left( 1 + k_1 - \frac{k_1}{p_1}\right)
\left( 1 + k_2 - \frac{k_2}{p_2}\right)\cdot \dots
\cdot \left( 1 + k_n - \frac{k_n}{p_n}\right)N - 2N + 1 \ , \label{Slom} \end{equation}

\noindent where it was assumed that the prime number decomposition of
$N$ is $N = p_1^{k_1}\cdot p_2^{k_2} \cdot \dots \cdot p_n^{k_n}$ \cite{ZT}.

\

\noindent {\it Affine families}: A family of Hadamard matrices stemming
from a Hadamard matrix $H$ is said to be affine \cite{TZ} if it can be
written in the form

\begin{equation} H(a,b,\dots ) = H\circ \mbox{EXP}(iR) \ , \label{Raffin}
\end{equation}

\noindent where the product is the entrywise Hadamard product, the exponentiation
is entrywise too, and $R$ belongs to a subspace of the set of all real $N\times N$
matrices. The parameters $a,b,\dots $ parametrise that linear subspace. The
family is given in dephased form if $H$ is dephased and $R$ has only zeroes in
the first row and
column. The affine family itself has the topology of a multi-dimensional real
torus.

An example for $N = 6$ is \cite{Haagerup}

\begin{equation} F(a,b) = \frac{1}{\sqrt{6}}
\left( \begin{array}{cccccc} 1 & 1 & 1 & 1 & 1 & 1 \\
1 & \omega e^{ia} & \omega^2e^{ib} & \omega^3 & \omega^4e^{ia} & \omega^5e^{ib} \\
1 & \omega^2 & \omega^4 & 1 & \omega^2 & \omega^4 \\
1 & \omega^3e^{ia} & e^{ib} & \omega^3 & e^{ia} & \omega^3e^{ib} \\
1 & \omega^4 & \omega^2 & 1 & \omega^4 & \omega^2 \\
1 & \omega^5e^{ia} & \omega^4e^{ib} & \omega^3 & \omega^2e^{ia} & \omega e^{ib}
\end{array}\right) \ , \hspace{8mm} \omega =e^{\frac{2\pi i}{6}} \ .
\label{Fourier6}
\end{equation}

\noindent Here $a$ and $b$ are phases that can be chosen freely. The corresponding
matrix $X$ is given to first order in eq. (\ref{N=6}). Taking the
transpose of the matrices in $F(a,b)$ yields another affine family
intersecting the first in the Fourier matrix.

\

\noindent {\it The Di\c{t}\u{a} construction}:
There exists a construction due to Di\c{t}\u{a} \cite{Dita} allowing us to
construct an affine family in dimension $N = N_1N_2$ starting from one
Hadamard matrix $H^{(0)}$ in dimension $N_1$ and $N_1$ possibly different
Hadamard matrices $H^{(1)}, \dots , H^{(N_1)}$ in dimension $N_2$. In dephased
form

\begin{equation} H = \left( \begin{array}{cccc} H^{(0)}_{0,0} H^{(1)} & H^{(0)}_{0,1}
D^{(1)}H^{(2)} & \dots & H^{(0)}_{0,N_1-1}D^{(N_1-1)}H^{(N_1)} \\ \vdots & \vdots & & \vdots \\
H^{(0)}_{N_1-1,0} H^{(1)} & H^{(0)}_{N_1-1,1}
D^{(1)}H^{(2)} & \dots & H^{(0)}_{N_1-1,N_1-1}D^{(N_1-1)}H^{(N_1)} \end{array}
\right) \label{Petre} \end{equation}

\noindent where $D^{(1)}, \dots , D^{(N_1-1)}$ are diagonal unitary matrices (with
their first entries equal to one in order to obtain $H$ in dephased form).

Now let $N = p_1^{k_1}\cdot p_2^{k_2}\cdot \dots \cdot p_n^{k_n}$ be the prime number
decomposition of $N$. Using the Fourier matrices $F_{p_1}, \dots , F_{p_n}$ as
seeds in the Di\c{t}\u{a} construction it is easy to show that we obtain an
affine family of dimension

\begin{equation} d_A = (k_1 + \dots + k_n -1)N - k_1\frac{N}{p_1} - \dots
- k_n\frac{N}{p_n} + 1 \ . \label{dA} \end{equation}

\noindent This is the dimension of the largest affine family obtainable in this way. 
But does it contain the Fourier matrix $F_N$?

Consider an affine family in dimension $N = N_1N_2$ obtained from the Di\c{t}\u{a} construction by setting $H^{(0)} = F_{N_1}$, $H^{(1)}= ... = H^{(N_1)} = F_{N_2}$ in
eq. (\ref{Petre}). If the parameters in the diagonal matrices $D^{(k)}$ are chosen
so that they become identity matrices we obtain the matrix $F_{N_1}\otimes F_{N_2}$.
But this family also contains a matrix equivalent to $F_N$. To see this, let

\begin{equation} 0 \leq r,s < N_1 \ , \hspace{10mm} 0 \leq m,n < N_2 \ ,
\end{equation}

\begin{equation} \omega = e^{\frac{2\pi i}{N}} \ , \hspace{5mm} q_1 =
\omega^{N_2} \ , \hspace{5mm} q_2 = \omega^{N_1} \ , \end{equation}

\noindent and introduce $N_1-1$ diagonal unitaries

\begin{equation} D^{(r)} = \mbox{diag}(1,x_1^{(r)}, \dots , x_{N_2-1}^{(r)}) \ .
\end{equation}

\noindent The Di\c{t}\u{a} construction gives rise to a Hadamard matrix
$H$ with matrix elements

\begin{equation} H_{rN_2+m, sN_2+n} = \frac{1}{\sqrt{N}}x_m^{(s)}q_1^{rs}q_2^{mn}
\ . \end{equation}

\noindent We now perform a column permutation and obtain an equivalent Hadamard
matrix $H'$ with matrix elements

\begin{equation} H'_{rN_2+m, nN_1+s} = \frac{1}{\sqrt{N}}x_m^{(s)}q_1^{rs}q_2^{mn}
\ . \label{N1N2}
\end{equation}

\noindent The Fourier matrix has the elements

\begin{equation} \frac{1}{\sqrt{N}}\omega^{ij} =
\frac{1}{\sqrt{N}}\omega^{(rN_2+m)(nN_1+s)} = \frac{1}{\sqrt{N}}\omega^{ms}
q_1^{rs}q_2^{mn} \ , \end{equation}

\noindent and is obtained from $H'$ by setting

\begin{equation} x_m^{(s)} = \omega^{ms} \ . \end{equation}

\noindent In prime power dimensions the affine family interpolates between the
non-equivalent matrices $F_N$ and $F_p\otimes \dots \otimes F_p$.

For $N = p^k$ the dimension $d_A$ equals the linear defect $d_1$ of the Fourier
matrix, so affine families
of larger dimensions cannot contain it. We believe that this is so for
all $N$; it is known to be so for $N \leq 18$ \cite{TZ,Wpers}. Affine families not
including the Fourier matrix, and not obtainable from the Di\c{t}\u{a} construction,
are known \cite{Szoll}.

\

\noindent {\it Self-cognate affine families}: By definition \cite{TZ} a self-cognate
family of Hadamard matrices goes into itself under transposition. In section
9 we use the fact that self-cognate affine families of dimension $d_A$ exist
whenever $N = p_1p_2^2$. To prove this let

\begin{equation} 0 \leq r,s < p_1 \ , \hspace{10mm} 0 \leq m,n , u, v < p_2 \ .
\end{equation}

\noindent We assume that the Di\c{t}\u{a} construction has already been applied
to construct a $p_1p_2 \times p_1p_2$ Hadamard matrix, as in eq.
(\ref{N1N2}). In the next step we use $p_2$ matrices of this type, and moreover the
diagonal unitaries used in the first step are allowed to differ from each other.
Thus there are
$(p_1-1)p_2$ diagonal unitaries $D^{(s,v)}$ from the first step, and an
additional $p_2-1$ diagonal unitaries

\begin{equation} d^{(v)} = \mbox{diag}(1,y_1^{(v)}, \dots , y_{p_1p_2-1}^{(v)}) \ .
\end{equation}

\noindent Using $H^{(0)} = F_{p_2}$, with matrix elements $q_2^{uv}/\sqrt{p_2}$,
the Di\c{t}\u{a} construction now gives a Hadamard matrix with matrix
elements

\begin{equation} H_{p_1p_2u+rp_2+m, p_1p_2v+np_1+s} = \frac{1}{\sqrt{N}}
q_1^{rs}q_2^{mn+uv}x_m^{(s,v)}y_{rp_2+m}^{(v)} \ . \end{equation}

\noindent We next perform a column permutation according to

\begin{equation} H'_{p_1p_2u+rp_2+m, p_1p_2n+sp_2+v} = \frac{1}{\sqrt{N}}
q_1^{rs}q_2^{mn+uv}x_m^{(s,v)}y_{rp_2+m}^{(v)} \end{equation}

\noindent These are the matrix elements of the matrix $H'(x,y)$. We want to
prove that for all choices of the parameters $x,y$ we
can find some parameters $\tilde{x}, \tilde{y}$ such that $H'(x,y)^{\rm T}
= H'(\tilde{x}, \tilde{y})$. Transposition of $H'$ is given by

\begin{equation} u \leftrightarrow n \ , \hspace{5mm} r \leftrightarrow s \ ,
\hspace{5mm} m \leftrightarrow v \ , \end{equation}

\noindent and by

\begin{equation} x_m^{(s,v)} \rightarrow x_v^{(r,m)} \ , \hspace{10mm}
y_{rp_2+m}^{(v)} \rightarrow y_{sp_2+v}^{(m)} \ . \end{equation}

\noindent So we set

\begin{equation} \tilde{x}^{(s,v)}_m = y_{sp_2+v}^{(m)} \ , \hspace{8mm}
\tilde{y}_{rp_2+m}^{(v)} = x_v^{(r,m)} \ , \end{equation}

\noindent which is always possible. This proves that the family is
self-cognate.

\

\noindent {\it Other affine families}: One obtains affine families
that do not go into themselves under transposition if one performs
the Di\c{t}\u{a} construction in a different order. The self-cognate
family was obtained by starting with a matrix of size $p_2$, enlarging
it to size $p_1p_2$, and finally to size $p_2p_1p_2$. Other examples
referred to in section 9 are obtained from the sequences $p_2 \rightarrow
p_2^2 \rightarrow p_1p_2^2$ and $p_1 \rightarrow p_2p_1 \rightarrow
p_2^2p_1$, and they lead to affine families of the same dimension. In the
special case of $N = 12$ all affine families of maximal dimension are known
\cite{TZ}, and we have proved that all of them can be obtained from some
variant of the Di\c{t}\u{a} construction.

\

\noindent {\it The case $N = 6$}: A number of non-affine families
of $6\times 6$ Hadamard matrices have been found \cite{Nicoara, Matolcsi,
Feri, Karlsson2}, and all analytically
known families are included as subfamilies of the
3-dimensional non-affine family found by Karlsson \cite{Karlsson}. As mentioned
in the introduction it is known that a 4-dimensional family exists
\cite{Feriphd}, but a proof that this family includes the Fourier matrix
is missing. It is also known that at least one isolated Hadamard
matrix, not belonging to this family, exists \cite{Butson, Tao}.

\

\noindent {\it Families not including $F_N$}: Finally, to avoid any misunderstanding,
we observe that although the defect
vanishes for $N = 7$ (say), this does not mean that the Fourier matrix is
the only known Hadamard matrix in this case. Indeed a one-dimensional affine 
family not including the Fourier matrix is known for $N = 7$, and for some
other cases where $N$ is a prime equal to 1 modulo 6 \cite{Petrescu}.
We have nothing to say about such families.

\vspace{10mm}

{\bf Appendix B: The prime power case}

\vspace{5mm}

\noindent When $N = p^k$ is a power of a prime $p$ the Di\c{t}\u{a} construction allows
us to construct an affine family of maximal dimension equal to the linear
defect $d_1$. A more direct construction is the following. The matrix elements
of $R$, in eq. (\ref{Raffin}), are
given in terms of independent parameters $\phi_{n,i,j}$ as

\begin{equation} R_{ij} = \sum_{n=0}^{k}\phi_{n,i \ {\rm mod} \ p^n,
j \ {\rm mod} \ p^{k-n}} \ , \hspace{10mm} \phi_{n,i,j} = 0 \ \  \mbox{if}
\ \ i < p^{n-1} \ . \end{equation}

\noindent Dephased matrices are obtained by excluding $n = 0$ and $n = k$ from
the sum, and setting $\phi_{n,i,0} = 0$, that is

\begin{equation} R_{ij} = \sum_{n=1}^{k-1}\phi_{n,i \ {\rm mod} \ p^n,
j \ {\rm mod} \ p^{k-n}} \ , \hspace{6mm} \phi_{n,i,j} = 0 \ \  \mbox{if}
\ \ i < p^{n-1} \ \mbox{or} \ j = 0 \ . \end{equation}

\noindent The modular arithmetic means that $R$ splits into $p\times p$ equal
blocks of size $p^{k-1}$. It is obtained by adding the matrices
$\phi_n$ together, and the number of entries in these matrices taken
together is

\begin{equation} \sum_{n=1}^{k-1}(p^n-p^{n-1})(p^{k-n}-1) = (k-1)p^k -
kp^{k-1} + 1 = d_1 \ . \end{equation}

\noindent Since $\phi_{n,i,j}$ repeats for $i \geq p^n$ while
$\phi_{n+1,i,j} = 0$ for $i < p^n$, these entries are indeed the independent
parameters on which the matrix $R$ depends. The proof that the resulting
matrices are unitary
Hadamard matrices is not entirely straightforward, but we omit it here.

The connection to the parametrisation we use in the main body of the paper is
easily found for the special case $N = p^2$. We simply compute $X = {\bf 1} - HF^\dagger$, 
and find for its matrix elements that 

\begin{equation} X_{i \ {\rm mod} \ p,rp \ {\rm mod} \ N} = - \frac{1}{p}\sum_{k=0}^{p-1}\omega^{-pkr}
e^{i\phi_{1,i \ {\rm mod} \ p,k \ {\rm mod} \ p}} \ . \end{equation}

\noindent From this we learn that the choice of free parameters that we make
in the perturbative construction may not be the optimal one for a closed form
expression.

\vspace{10mm}

{\bf Appendix C: The consistency conditions hold to second order}

\vspace{5mm}

\noindent We mention in the text that the consistency conditions (\ref{cons})
always hold to second order. We give the key steps in the proof here since
they serve to illustrate the complexities involved in making a
calculation valid for all $N$. In particular we need to make a liberal use
of linear congruence relations.

Setting $s = 2$ in eq. (\ref{Bsn}) and replacing $X^{(1)}$ by its solution
(\ref{losning}) we obtain

\begin{eqnarray} B_{i,i}^{(2,n)} = \hspace{70mm} \nonumber \\ \ \\
= \sum_{k=0}^{N-1}\left[ x^{(1)}_{(i+n)
\ {\rm mod} \ {\rm gcd}(k-n,N),k-n} - x^{(1)}_{i \ {\rm mod} \ {\rm gcd} (k-n,N),
k-n}\right] x^{(1)}_{i \ {\rm mod} \ {\rm gcd}(k,N),-k} \ . \nonumber
\end{eqnarray}

\noindent At second order therefore the consistency conditions (\ref{cons}) are

\begin{eqnarray} 0 = \sum_{q=0}^{\frac{N}{{\rm gcd}(n,N)} - 1}\sum_{k=0}^{N-1}
x^{(1)}_{(i+q \ {\rm gcd}(n,N)) \ {\rm mod \ gcd}(k,N),-k} \times
\hspace{25mm} \nonumber \\
\ \\ \times \left[
x^{(1)}_{(i+q \ {\rm gcd}(n,N) + n) \ {\rm mod} \ {\rm gcd}(k-n,N),k-n} -
x^{(1)}_{(i+q \ {\rm gcd}(n,N)) \ {\rm mod} \ {\rm gcd}(k-n,N),k-n} \right]
\nonumber \ .
\end{eqnarray}

\noindent The crucial step is to realise that the first factor is the same
for two values of $q$ whenever their difference $\Delta q$ obeys

\begin{equation} \Delta q \ \mbox{gcd}(n,N) = 0 \ \ \mbox{mod gcd}(k,N) \ .
\end{equation}

\noindent This will permit us to break the sum over $q$ into two, one of
which involves only the second factor.

The smallest solution for $\Delta q$ is

\begin{equation} \Delta q = \frac{\mbox{lcm}\left( \mbox{gcd}(n,N),\mbox{gcd}
(k,N)\right)}{\mbox{gcd}(n,N)} = \frac{\mbox{gcd}\left( \mbox{lcm}(n,k),N\right)}
{\mbox{gcd}(n,N)} \ , \end{equation}

\noindent where lcm denotes the least common multiple. Write
$q = q_1 + \Delta q \ q_2$. The consistency condition then takes the form

\begin{eqnarray} 0 = \sum_{k=0}^{N-1}\sum_{q_1=0}^{\frac{{\rm gcd}({\rm lcm}(n,k),N))}{{\rm gcd}(n,N)} - 1} x^{(1)}_{(i+q_1 \ {\rm gcd}(n,N)) \ {\rm mod \ gcd}(k,N),-k} \times
\hspace{15mm} \nonumber \\ \label{slutet} \\
\times \sum_{q_2 = 0}^{\frac{N}{{\rm gcd}({\rm lcm}(n,k),N)}-1}
\Large[ x^{(1)}_{(i+q_1 \ {\rm gcd}(n,N) + q_2 \ {\rm gcd}({\rm lcm}(n,k),N) + n)
\ {\rm mod} \ {\rm gcd}(k-n,N),k-n} - \nonumber \\
\nonumber \\
- x^{(1)}_{(i+q_1 \ {\rm gcd}(n,N) + q_2 \ {\rm gcd}({\rm lcm}(n,k),N)) \
{\rm mod} \ {\rm gcd}(k-n,N),k-n} \Large] \ .
\nonumber \end{eqnarray}

\noindent The dependence on $q_2$ is now isolated to the sum that constitutes
the second factor, and it is enough to show that this sum vanishes. Indeed the terms
will cancel in pairs if
for each $q_2$ we can find a $q_2 + \Delta q_2$ such that

\begin{equation} \left\{ \Delta q_2 \ \mbox{gcd}({\rm lcm}(n,k),N) = n\right\}
\ \ \mbox{mod gcd}(k-n,N) \ . \end{equation}

\noindent This is what we need, because although $q_2$ and $\Delta q_2$ are defined
modulo $N/{\rm gcd}({\rm lcm}(n,k),N)$ they appear multiplied by
${\rm gcd}({\rm lcm}(n,k),N)$ in eq. (\ref{slutet}). We know that the equation
$xa = y$ taken modulo $b$ has a solution if and only
if gcd$(a,b)$ divides $y$. So we must show that $n$ is divisible by

\begin{equation} \mbox{gcd}\left( \mbox{gcd}(\mbox{lcm}(n,k),N), \mbox{gcd}(k-n,N)\right)
= \mbox{gcd}\left( \mbox{lcm}(n,k),k-n,N)\right) \ . \end{equation}

\noindent It is enough to show that $n$ is divisible by
$\mbox{gcd}\left( \mbox{lcm}(n,k),k-n\right)$. But this is so because

\begin{equation} \mbox{gcd}\left( \mbox{lcm}(n,k),k-n\right) = \mbox{lcm} \left(
\mbox{gcd}(n,k-n), \mbox{gcd}(k,k-n)\right) = \mbox{gcd}(n,k) \ \end{equation}

\noindent which certainly divides $n$. This ends the proof.

\vspace{10mm}

{\bf Appendix D: A toy model}

\vspace{5mm}

\noindent The description of our method in section 2 may appear forbidding, so in 
this Appendix we apply it to the simple example 

\begin{equation} f(X,Y) = X\left( X-1\right) ^{2}- \left( {\rm e}^{Y}-1\right)^{2} = 0 \ . 
\label{blo0}
\end{equation}

\noindent Since there are only two variables we streamline the notation a little. 
The two solutions $\left( X,Y\right) =\left( 0,0\right) $ and
$\left( X,Y\right) =\left( 1,0\right) $ are obvious by inspection. We shall 
work up to order $4$. 

For the first solution we expand eq. (\ref{blo0}) around the origin:

\begin{equation}
f(X,Y) = X-2X^{2}+X^{3}-Y^{2}-Y^{3}-7Y^{4}/12 + \dots \ .  \label{blo2}
\end{equation}

\noindent We next collect $X$ and $Y$ into a vector and expand it as in eq. (\ref{expand}),

\begin{equation}
\left[
\begin{array}{c}
X \\
Y
\end{array}
\right] =\left[
\begin{array}{c}
X_{\left( 1\right) } \\
Y_{\left( 1\right) }
\end{array}
\right] +\left[
\begin{array}{c}
X_{\left( 2\right) } \\
Y_{\left( 2\right) }
\end{array}
\right] +\left[
\begin{array}{c}
X_{\left( 3\right) } \\
Y_{\left( 3\right) }
\end{array}
\right] +\left[
\begin{array}{c}
X_{\left( 4\right) } \\
X_{\left( 4\right) }
\end{array}
\right] + \dots \ .  \label{blo1}
\end{equation}

\noindent We then obtain eqs. (9-12) in the form  

\begin{eqnarray}
X_{\left( 1\right) } &=&0 \hspace{68mm} \equiv B_{(1)} \\
X_{\left( 2\right) } &=&2X_{\left( 1\right) }^{2}+Y_{\left( 1\right) }^{2} 
\hspace{49mm} \equiv B_{\left( 2\right) } \\
X_{\left( 3\right) } &=&-X_{\left( 1\right) }^{3}+4X_{\left( 1\right)
}X_{\left( 2\right) }+Y_{\left( 1\right) }^{3}+2Y_{\left( 1\right)
}Y_{\left( 2\right) } \hspace{6mm} \equiv B_{\left( 3\right) } \\
X_{\left( 4\right) } &=&2X_{\left( 2\right) }^{2}+4X_{\left( 1\right)
}X_{\left( 3\right) }-3X_{\left( 1\right) }^{2}X_{\left( 2\right)
}+Y_{\left( 2\right) }^{2}+  \nonumber \\
&&+2Y_{\left( 1\right) }Y_{\left( 3\right) }+3Y_{\left( 1\right)
}^{2}Y_{\left( 2\right) }+7Y_{\left( 1\right) }^{4}/12 \hspace{12mm} 
\equiv B_{\left( 4\right) } \ . 
\end{eqnarray}

\noindent At each order this is the linear system $AX_{(n)} = B_{(n)}$, where
$A = \left[ \begin{array}{cc} 1 & 0 \end{array} \right]$. 
The linear defect is the number of variables minus the rank of $A$, and equals $1$.

We choose the Moore-Penrose pseudo-inverse 

\begin{equation}
\hat{A}=\left[
\begin{array}{c}
1 \\
0
\end{array}
\right] \quad ,\quad 1-\hat{A}A=\left[
\begin{array}{cc}
0 & 0 \\
0 & 1
\end{array}
\right] \quad ,\quad 1-A\hat{A}=0
\ . \end{equation}

\noindent In our solution (15) we denote the components of the arbitrary vector $Z$ 
by $x_{(n)}$ and $y_{(n)}$, and obtain at each order 

\begin{equation}
\left[
\begin{array}{c}
X_{\left( n\right) } \\
Y_{\left( n\right) }
\end{array}
\right] =\left[
\begin{array}{c}
1 \\
0
\end{array}
\right] \left[ B_{\left( n\right) }\right] +\left[
\begin{array}{cc}
0 & 0 \\
0 & 1
\end{array}
\right] \left[
\begin{array}{c}
x_{\left( n\right) } \\
y_{\left( n\right) }
\end{array}
\right] =\left[
\begin{array}{c}
B_{\left( n\right) } \\
y_{\left( n\right) }
\end{array}
\right] \ . 
\end{equation}

\noindent There are no consistency conditions, which means that the defect remains $1$
to all orders. Hence the solution is one-dimensional and unique around this
point. 

We have $B_{(1)} = 0$, and the remaining $B_{(n)}$ are computed recursively. 
To order 4 we obtain  

\begin{eqnarray}
\left[
\begin{array}{c}
X \\
Y
\end{array}
\right]  &=&\left[
\begin{array}{c}
\left\{
\begin{array}{c}
y_{\left( 1\right) }^{2}+y_{\left( 1\right) }^{3}+2y_{\left( 1\right)
}y_{\left( 2\right) }+y_{\left( 2\right) }^{2}+ \\
+2y_{\left( 1\right) }y_{\left( 3\right) }+3y_{\left( 1\right)
}^{2}y_{\left( 2\right) }+31y_{\left( 1\right) }^{4}/12
\end{array}
\right\}  \\
y_{\left( 1\right) }+y_{\left( 2\right) }+y_{\left( 3\right) }+y_{\left(
4\right) }
\end{array}
\right] \ . \end{eqnarray} 

\noindent Setting $t = y_{\left( 1\right) }+y_{\left( 2\right) }+
y_{\left( 3\right) }+y_{\left(4\right) }$ our first solution is, to order 4,

\begin{eqnarray} \left[
\begin{array}{c}
X \\
Y
\end{array}
\right]  &\approx & \left[
\begin{array}{c} t^2 + t^3 + 31 t^4/12 \\ t 
\end{array}
\right] \ .   \label{so1}
\end{eqnarray}

Next we expand around the solution $(X,Y) = (1,0)$. After shifting the variable 
$X$, and expanding around the new origin, we get  

\begin{equation}
f(X,Y) = X^{2}+X^{3}-Y^{2}-Y^{3}-7Y^{4}/12 + \dots = 0 \ . \label{blo3}
\end{equation}

\noindent Again we expand $X$ and $Y$ as in eq. (\ref{expand}) and obtain eqs. 
(9-12) in the form  

\begin{eqnarray}
0 &=&0 \hspace{70mm} \equiv B_{(1)} \\
0 &=&-X_{\left( 1\right) }^{2}+Y_{\left( 1\right) }^{2} \hspace{50mm} 
\equiv B_{(2)} \\
0 &=&-X_{\left( 1\right) }^{3}-2X_{\left( 1\right) }X_{\left( 2\right)
}+Y_{\left( 1\right) }^{3}+2Y_{\left( 1\right) }Y_{\left( 2\right) } \hspace{8mm} 
\equiv B_{(3)} \\
0 &=&-X_{\left( 2\right) }^{2}-2X_{\left( 1\right) }X_{\left( 3\right)
}-3X_{\left( 1\right) }^{2}X_{\left( 2\right) }+Y_{\left( 2\right) }^{2}+
\nonumber \\
&&+2Y_{\left( 1\right) }Y_{\left( 3\right) }+3Y_{\left( 1\right)
}^{2}Y_{\left( 2\right) }+7Y_{\left( 1\right) }^{4}/12 \hspace{14mm} \equiv B_{(4)} \ . 
\end{eqnarray}

\noindent We have obtained the linear systems $AX_{(n)} = B_{(n)}$, where $A = \left[
\begin{array}{cc} 0 & 0 \end{array} \right]$.
Again we use the Moore-Penrose inverse 

\begin{equation}
\hat{A}=\left[
\begin{array}{c}
0 \\
0
\end{array}
\right] \quad ,\quad 1-\hat{A}A=\left[
\begin{array}{cc}
1 & 0 \\
0 & 1
\end{array}
\right] \quad ,\quad 1-A\hat{A}=1 \ . 
\end{equation}

\noindent Our solution (15) becomes

\begin{equation}
\left[
\begin{array}{c}
X_{\left( n\right) } \\
Y_{\left( n\right) }
\end{array}
\right] =\left[
\begin{array}{c}
0 \\
0
\end{array}
\right] \left[ B_{\left( n\right) }\right] +\left[
\begin{array}{cc}
1 & 0 \\
0 & 1
\end{array}
\right] \left[
\begin{array}{c}
x_{\left( n\right) } \\
y_{\left( n\right) }
\end{array}
\right] =\left[
\begin{array}{c}
x_{\left( n\right) } \\
y_{\left( n\right) }
\end{array}
\right]  \ , \label{blo4}
\end{equation}

\noindent subject to the consistency condition (14), which takes the form 
\begin{equation}
B_{\left( n\right) }=0
\end{equation}

\noindent The linear defect is $2$. Since the solutions (\ref{blo4}) have no heterogeneous
component, all we have to do is to study the consistency conditions. To
second order this is
\begin{equation}
B_{\left( 2\right) }=-X_{\left( 1\right) }^{2}+Y_{\left( 1\right)
}^{2}=-x_{\left( 1\right) }^{2}+y_{\left( 1\right) }^{2}=0 \ , \label{blo5}
\end{equation}
with two solutions
\begin{eqnarray}
\mathrm{I.}\quad y_{\left( 1\right) } &=&x_{\left( 1\right) } \\
\mathrm{II.}\quad y_{\left( 1\right) } &=&-x_{\left( 1\right) }
\end{eqnarray}
Therefore the second order defect is $1$.

To third order we have

\begin{eqnarray}
B_{\left( 3\right) } &=&-X_{\left( 1\right) }^{3}-2X_{\left( 1\right)
}X_{\left( 2\right) }+Y_{\left( 1\right) }^{3}+2Y_{\left( 1\right)
}Y_{\left( 2\right) }=  \nonumber \\
&=&-x_{\left( 1\right) }^{3}-2x_{\left( 1\right) }x_{\left( 2\right)
}+y_{\left( 1\right) }^{3}+2y_{\left( 1\right) }y_{\left( 2\right) }=0 \ . 
\label{blo6}
\end{eqnarray}

\noindent For solution I this is

\begin{equation}
B_{\left( 3\right) }=\left[
\begin{array}{cc}
-2x_{\left( 1\right) } & 2x_{\left( 1\right) }
\end{array}
\right] \left[
\begin{array}{c}
x_{\left( 2\right) } \\
y_{\left( 2\right) }
\end{array}
\right] =0\quad \Rightarrow \quad y_{\left( 2\right) }=x_{\left( 2\right) } \ . 
\end{equation}

\noindent This is the unnumbered equation $Uh_{n}^{\left( 2\right) }=V$ mentioned 
at the end of section 2, with $U=\left[
\begin{array}{cc}
-2x_{\left( 1\right) } & 2x_{\left( 1\right) }
\end{array}
\right] $ and $V=0$. For solution II we get

\begin{eqnarray}
B_{\left( 3\right) } &=&\left[
\begin{array}{cc}
-2x_{\left( 1\right) } & -2x_{\left( 1\right) }
\end{array}
\right] \left[
\begin{array}{c}
x_{\left( 2\right) } \\
y_{\left( 2\right) }
\end{array}
\right] -2x_{\left( 1\right) }^{3}=0 \\ 
&\Rightarrow &\quad y_{\left( 2\right) }=-x_{\left( 2\right) }-x_{\left(
1\right) }^{2} \ . 
\end{eqnarray}

\noindent Here $U=\left[
\begin{array}{cc}
-2x_{\left( 1\right) } & 2x_{\left( 1\right) }
\end{array}
\right] $ and $V=2x_{\left( 1\right) }^{3}$.
At second order we fixed the variable $y_{\left( 1\right) }$ and 
at third order we fixed $y_{(2)}$. 
No further fixing of first order variables was
needed, so the defect remains $1$. 

Note that $x_{\left( 1\right)
}=y_{\left( 1\right) }=0$ satisfies both (\ref{blo5}) and (\ref{blo6}). A 
further analysis reveals that this leads to special cases of solutions I and II.

To fourth order we have

\begin{eqnarray}
B_{\left( 4\right) }&=&-x_{\left( 2\right) }^{2}-2x_{\left( 1\right) }x_{\left(
3\right) }-3x_{\left( 1\right) }^{2}x_{\left( 2\right) }+
\nonumber \\
&&+y_{\left( 2\right) }^{2}+2y_{\left( 1\right) }y_{\left( 3\right)
}+3y_{\left( 1\right) }^{2}y_{\left( 2\right) }+7y_{\left( 1\right)
}^{4}/12=0 \label{blo7}
\end{eqnarray}

\noindent For solution I we obtain an equation $Uh_{n}^{\left( 3\right) }=V$ of the form 
\begin{eqnarray}
B_{\left( 4\right) } &=&\left[
\begin{array}{cc}
-2x_{\left( 1\right) } & 2x_{\left( 1\right) }
\end{array}
\right] \left[
\begin{array}{c}
x_{\left( 3\right) } \\
y_{\left( 3\right) }
\end{array}
\right] +7x_{\left( 1\right) }^{4}/12=0 \\ 
&\Rightarrow &\quad y_{\left( 3\right) }=x_{\left( 3\right) }-7x_{\left(
1\right) }^{3}/24
\end{eqnarray}

\noindent The same matrix $U=\left[
\begin{array}{cc}
-2x_{\left( 1\right) } & 2x_{\left( 1\right) }
\end{array}
\right] $ shows up again, while now $V=-7x_{\left( 1\right) }^{4}/12$. And for
solution II we get

\begin{eqnarray}
B_{\left( 4\right) } &=&\left[
\begin{array}{cc}
-2x_{\left( 1\right) } & -2x_{\left( 1\right) }
\end{array}
\right] \left[
\begin{array}{c}
x_{\left( 3\right) } \\
y_{\left( 3\right) }
\end{array}
\right] -4x_{\left( 1\right) }^{2}x_{\left( 2\right) }-17x_{\left( 1\right)
}^{4}/12=0 \\ 
&\Rightarrow &\quad y_{\left( 3\right) }=-x_{\left( 3\right) }-2x_{\left(
1\right) }x_{\left( 2\right) }-17x_{\left( 1\right) }^{3}/24
\end{eqnarray}

\noindent Again the same matrix $U$ shows up.

And again at this order $y_{\left( 3\right) }$ is fixed and no further
fixing of first order variables is needed, so the defect remains $1$.
Summing up the order-by-order solutions one gets the two solutions
\begin{eqnarray}
&&\mathrm{I.}\quad \left[
\begin{array}{c}
X \\
Y
\end{array}
\right] =\left[
\begin{array}{c}
x_{\left( 1\right) }+x_{\left( 2\right) }+x_{\left( 3\right) }+x_{\left(
4\right) } \\
x_{\left( 1\right) }+x_{\left( 2\right) }+x_{\left( 3\right) }-7x_{\left(
1\right) }^{3}/24+y_{\left( 4\right) }
\end{array}
\right] \approx  \\
&\approx &\left[
\begin{array}{c}
x_{\left( 1\right) }+x_{\left( 2\right) }+x_{\left( 3\right) } \\
\left( x_{\left( 1\right) }+x_{\left( 2\right) }+x_{\left( 3\right) }\right)
-7\left( x_{\left( 1\right) }+x_{\left( 2\right) }+x_{\left( 3\right)
}\right) ^{3}/24
\end{array}
\right] +\left[
\begin{array}{c}
x_{\left( 4\right) } \\
y_{\left( 4\right) }
\end{array}
\right]   \label{so2} \\
&&\mathrm{II.}\quad \left[
\begin{array}{c}
X \\
Y
\end{array}
\right] =\left[
\begin{array}{c}
x_{\left( 1\right) }+x_{\left( 2\right) }+x_{\left( 3\right) }+x_{\left(
4\right) } \\
\left\{
\begin{array}{c}
-x_{\left( 1\right) }-x_{\left( 2\right) }-x_{\left( 1\right)
}^{2}-x_{\left( 3\right) }- \\
-2x_{\left( 1\right) }x_{\left( 2\right) }-17x_{\left( 1\right)
}^{3}/24+y_{\left( 4\right) }
\end{array}
\right\}
\end{array}
\right] \approx  \\
&\approx &\left[
\begin{array}{c}
x_{\left( 1\right) }+x_{\left( 2\right) }+x_{\left( 3\right) } \\
\left\{
\begin{array}{c}
-\left( x_{\left( 1\right) }+x_{\left( 2\right) }+x_{\left( 3\right)
}\right) -\left( x_{\left( 1\right) }+x_{\left( 2\right) }+\right.  \\
\left. +x_{\left( 3\right) }\right) ^{2}-17\left( x_{\left( 1\right)
}+x_{\left( 2\right) }+x_{\left( 3\right) }\right) ^{3}/24
\end{array}
\right\}
\end{array}
\right] +\left[
\begin{array}{c}
x_{\left( 4\right) } \\
y_{\left( 4\right) }
\end{array}
\right]   \label{so3}
\end{eqnarray}

\begin{figure}
        \centerline{ \hbox{
                \epsfig{figure=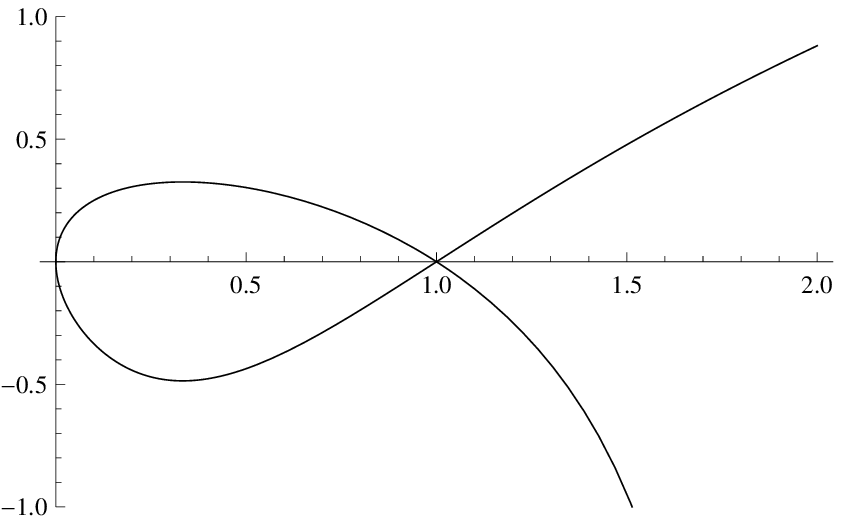,width=65mm}}}
        \caption{\small Solution to $X\left( X-1\right)^2=\left({\rm e}^Y-1\right)^2$.}
        \label{figura}
\end{figure}

We now compare our results with the known form of the solution, see Fig. \ref
{figura}. Around $\left( X,Y\right) =\left( 0,0\right) $ the solution is
smooth and unique. We can use the formula for the solution of the cubic and trigonometric
identities to write it as

\begin{equation}
X=\frac{4}{3}\sin ^{2}\left[ \frac{1}{3}\arcsin \left( \frac{3\sqrt{3}}{2}%
\left( e^{Y}-1\right) \right) \right] \approx Y^{2}+Y^{3}+\frac{31}{12}Y^{4} \ , 
\end{equation}

\noindent in agreement with eq. (\ref{so1}). At $\left( X,Y\right) =\left(
1,0\right) $ there are two intersecting solutions. Around
this point it is easy to solve eq. (\ref{blo0}) to get

\begin{eqnarray}
\mathrm{I.}\quad Y &=&\ln \left[ 1+\left( X-1\right) \sqrt{X}\right] \approx
\left( X-1\right) -\frac{7}{24}\left( X-1\right) ^{3} \\
\mathrm{II.}\quad Y &=&\ln \left[ 1-\left( X-1\right)
\sqrt{X}\right] \approx -\left( X-1\right) -\left( X-1\right) ^{2}-
\nonumber\\
&&-\frac{17}{24}\left( X-1\right) ^{3}
\end{eqnarray}

\noindent in agreement with eqs. (\ref{so2}) and (\ref{so3}) respectively (if we recall 
the shift we made in $X$).

One final word about the choice of the homogeneous terms and how it relates to different 
parametrizations of the solution. By setting all homogeneous
terms with $n>1$ to $0$, we are using $Y$ as the free parameter in (\ref{so1}%
) and $X$ in (\ref{so2}) and (\ref{so3}). But if instead we choose $x_{\left(
1\right) }=-t$, $x_{\left( 2\right) }=-t^{2}$, $x_{\left( 3\right)
}=-7t^{3}/24$ in eq. (\ref{so3}) we get 

\begin{equation}
\mathrm{II.}\quad \left[
\begin{array}{c}
X \\
Y
\end{array}
\right] \approx \left[
\begin{array}{c}
-t-t^{2}-7t^{3}/24 \\
t
\end{array}
\right] \ . 
\end{equation}

\noindent This amounts to using $Y$ as the free parameter. In (\ref{so1}) it is
clearly impossible to use $X$ as the free parameter (because the
curve is tangent to the $Y$ axis here) but one can set $y_{\left( 1\right)
}=t$, $y_{\left( 2\right) }=-t^{2}/2$, $y_{\left( 3\right) }=-2t^{3}/3$ and
generate the equally valid parametrization

\begin{equation}
\left[
\begin{array}{c}
X \\
Y
\end{array}
\right] \approx \left[
\begin{array}{c}
t^{2} \\
t-t^{2}/2-2t^{3}/3
\end{array}
\right] \ . 
\end{equation}

\end{document}